\documentclass{aastex63}

\received{October 20, 2022}
\revised{December 19, 2022}
\accepted{December 20, 2022}

\submitjournal{ApJ}

\shorttitle{Mass Distribution of Icy Fragments}
\shortauthors{Hasegawa et al.}

\begin{document}

\title{Collisional Growth and Fragmentation of Dust Aggregates. II. Mass Distribution of Icy Fragments}

\correspondingauthor{Yukihiko Hasegawa}
\email{yukih@ea.c.u-tokyo.ac.jp}

\author[0000-0001-7298-2144]{Yukihiko Hasegawa}
\affiliation{Graduate School of Arts \& Sciences, The University of Tokyo, 3-8-1, Komaba, Meguro, Tokyo 153-8902, Japan}

\author[0000-0001-9734-9601]{Takeru K. Suzuki}
\affiliation{Graduate School of Arts \& Sciences, The University of Tokyo, 3-8-1, Komaba, Meguro, Tokyo 153-8902, Japan}

\author[0000-0001-9659-658X]{Hidekazu Tanaka}
\affiliation{Astronomical Institute, Graduate School of Science, Tohoku University, 6-3, Aramaki, Aoba-ku, Sendai 980-8578, Japan}

\author[0000-0001-8808-2132]{Hiroshi Kobayashi}
\affiliation{Department of Physics, Nagoya University, Nagoya, Aichi 464-8602, Japan}

\author[0000-0002-6710-1768]{Koji Wada}
\affiliation{Planetary Exploration Research Center, Chiba Institute of Technology, 2-17-1, Tsudanuma, Narashino, Chiba 275-0016, Japan}





\begin{abstract}

By performing $N$-body simulations, we investigated fundamental processes of collisions between dust aggregates composed of submicron-sized icy dust monomers.
We examined the mass distribution of fragments in the collisional outcomes in a wide range of the mass ratio and the collision velocity between colliding dust aggregates.
We derived analytic expressions of the mass distribution of large remnants and small fragments by numerical fitting to the simulation results.
Our analytic formulae for masses of the large remnants can reproduce the contribution of mass transfer from a large target to a small projectile, which occurs for a mass ratio of $\gtrsim 3$ and is shown in a previous study (Hasegawa et al. 2021).
We found that the power-law index of the cumulative mass distribution of the small fragments is independent of the mass ratio and only weakly dependent on the collision velocity.
On the other hand, the mass fraction of fragments of individual dust monomers decreases with an increasing total mass of colliding aggregates for a fixed mass ratio.
This tendency implies that multiple hierarchical disruptive collisions (i.e., collisions between fragments, collisions between fragments of fragments) are required for producing a large amount of individual dust monomers via collisional fragmentation.
Our fragment model suggests that the total geometric cross section integrated over the fragments is estimated to be about the same order of the geometric cross section of the target.

\end{abstract}


\keywords{methods: numerical --- planets and satellites: formation --- protoplanetary disks}


\section{Introduction} \label{sec:int}

One of the key mechanisms in planet formation is the collisional growth and fragmentation of dust particles in protoplanetary disks (Nakagawa et al. 1981, 1986; Birnstiel et al. 2010; Okuzumi et al. 2012; Kataoka et al. 2013; Dr\c{a}\.{z}kowska \& Dullemond 2014; Krijt et al. 2015; Homma et al. 2019; Kobayashi \& Tanaka 2021).
The size distribution of dust grains, as well as their internal structures, is altered by collisions between dust grains and subsequent ejection of remnants (Lange \& Ahrens 1982; Takagi et al. 1984; Mizutani et al. 1990; Dominik \& Tielens 1997; Blum \& Wurm 2000; Wada et al. 2007, 2008, 2009, 2011, 2013; Suyama et al. 2008, 2012; Paszun \& Dominik 2009; Teiser \& Wurm 2009; G\"{u}ttler et al. 2010; Kobayashi \& Tanaka 2010; Gundlach et al. 2011; Ringl et al. 2012; Meru et al. 2013; Seizinger et al. 2013; Gundlach \& Blum 2015; Gunkelmann et al. 2016; Musiolik et al. 2016a, 2016b; Schr\"{a}pler et al. 2018, 2022; Hasegawa et al. 2021; Arakawa et al. 2022).
The evolution of the dust size distribution is an essential building block in the planet formation process.

In Hasegawa et al. (2021, hereafter Paper I), we carried out impact simulations to investigate the critical collisional fragmentation velocity of dust aggregates in a wide range of the initial mass ratio between colliding dust aggregates.
We showed that the critical collisional fragmentation velocity strongly depends on the mass ratio.
Offset collisions between unequal-mass dust aggregates cause the mass transfer from a larger target to a smaller projectile even for the lower collision velocity than the critical fragmentation velocity for the equal-mass collision, although the mass transfer rarely occurs in head-on collisions.
In collisions between dust aggregates composed of submicron-sized icy dust monomers, the minimum critical fragmentation velocity averaged over the impact parameter, $\approx 25 ~ \mathrm{m ~ s^{-1}}$, is obtained for the mass ratio of 3, which is about half of that obtained from equal-mass collisions (Wada et al. 2009; Paper I).
On the other hand, Paper I also suggested that the critical collisional fragmentation velocity could be higher than 100 $\mathrm{m ~ s^{-1}}$ for a large mass ratio of $\approx$ 100, because the larger target absorbs the impact energy of the smaller projectile and consequently the projectile is buried in the target (Wada et al. 2013).

These newly obtained results are expected to affect the evolution of the size and mass distributions of dust particles.
When mass is transferred from a larger target to a smaller projectile via an unequal-mass collision, the mass ratio of the two colliding bodies is reduced after the collision.
A possible consequence of such mass-transfer events is coexistence of multiple large dust aggregates with similar masses (Paper I).
If the effect of the mass transfer was not taken into account, only a single or limited number of large aggregates would grow unless the target was catastrophically fragmented (Krijt et al. 2015).
Besides the evolution of large objects, it is still uncertain how the mass transfer quantitatively affects the mass distribution of small fragments.
Thus, in order to clarify the growth of dust in protoplanetary disks, it is important to quantitatively determine the mass distribution of all fragments by explicitly taking into account the mass transfer from larger bodies to smaller bodies.

In this paper, we study the mass distribution of the fragments after the collisions of icy dust aggregates with various mass ratios.
Our purpose is to model the mass distribution of the fragments of our $N$-body simulations as functions of the collision velocity and the mass ratio between colliding dust aggregates and derive analytic formulae that can be easily prescribed in global models for the evolution of dust in protoplanetary disks.
We briefly describe our simulation method in Section \ref{sec:pro} and our procedure to analyze the numerical results in Section \ref{sec:mod}.
In Section \ref{sec:res}, main results are presented.
In Section \ref{sec:dis}, we discuss the geometric cross sections in comparison to previous study and the slopes of the mass distributions of the fragments.
A summary of this study is in Section \ref{sec:sum}.

\section{Numerical Procedures} \label{sec:pro}

We carry out three-dimensional numerical $N$-body simulations of collisions between two water-ice dust aggregates by using the numerical code developed in Wada et al. (2009).
Our simulation method is the same as the one used in Paper I; we numerically simulate the motions of all dust monomers constituting dust aggregates.
The model of the interaction between the two contact dust monomers is based on the Johnson, Kendall, and Roberts (JKR) theory (Johnson et al. 1971; Johnson 1987), and the details of the interaction model are described in Wada et al. (2007).
We assume that all dust monomers are spheres with radius of 0.1 \micron ~ made of water ice of an internal density of 1 $\mathrm{g ~ cm^{-3}}$ (Paper I).
Properties of water ice and the critical rolling displacement are the same values adopted in Paper I, except for the surface energy in Section \ref{sec:dis_mat_ice}, and are summarized in Table \ref{tab:properties}.
The energy for breaking a single contact between two dust monomers in the equilibrium position is given by $E_{\rm bre} = 6.1 \times 10^{-10} ~ \mathrm{erg}$ for the surface energy of 100 $\mathrm{mJ ~ m^{-2}}$ (Wada et al. 2007, 2013), and the characteristic break-up velocity is defined as $v_{\rm bre} \equiv (E_{\rm bre}/m_{\rm mon})^{1/2} = 3.8 ~ \mathrm{m ~ s^{-1}}$, where $m_{\rm mon} = 4.2 \times 10^{-15} ~ \mathrm{g}$ is the monomer mass.
We note that the ratio of monomer numbers constituting dust aggregates corresponds to their mass ratio, because we are considering equal-mass monomers, and that the mass of a dust aggregate, $M$, composed of a monomer number $N$ is obtained as $M = N m_{\rm mon}$.

\begin{table}
  \centering
  \caption{Material Properties of Dust Monomers}
  \label{tab:properties}
  \begin{tabular}{lll}
    \hline
    \hline
    Material properties & Values & Reference \\
    \hline
    Radius & 0.1 \micron & Wada et al. (2007) \\
    Internal density & 1 $\mathrm{g ~ cm^{-3}}$ & Wada et al. (2007) \\
    Surface energy & $100 ~ \mathrm{mJ ~ m^{-2}} = 100 ~ \mathrm{erg ~ cm^{-2}}$ & Wada et al. (2007) \\
    { } & $50 ~ \mathrm{mJ ~ m^{-2}}$ (Section \ref{sec:dis_mat_ice}) & { } \\
    Young's modulus & $7 ~ \mathrm{GPa} = 7 \times 10^{10}$ $\mathrm{dyn ~ cm^{-2}}$ & Wada et al. (2007) \\
    Poisson's ratio & 0.25 & Wada et al. (2007) \\
    Critical rolling displacement & 8 \r{A} & Wada et al. (2009) \\
    \hline
  \end{tabular}
\end{table}

During a two-body collision, we call the larger (smaller) dust aggregate the target (projectile).
In this paper, we utilize the data obtained by Paper I.
We adopted ballistic particle-cluster aggregation (BPCA) clusters in Paper I in order to directly compared the results to those obtained by Wada et al. (2009) and (2013), who also employed BPCA clusters.
Following these previous studies, we also adopt BPCA clusters without spinning motion for both the target and the projectile, which are randomly generated in each simulation.

Input parameters of our simulations are the number of dust monomers of the target, $N_{\rm tar}$, and the projectile, $N_{\rm pro}$, the collision velocity, $v_{\rm col}$, and the impact parameter, $b_{\rm off}$.
For a set of these parameters, we perform four simulation runs with different initial distributions of monomers that make up the target and the projectile.
The average of a variable $A$ taken over the four runs is denoted by $\bar{A}$.
In addition, considering the dependence of the impact parameter on the collision frequency, we represent the average of a variable $A$ weighted over the impact parameter as
\begin{equation}
  \langle A \rangle \equiv \displaystyle \frac{\displaystyle \int _0^{b_{\rm max}} 2 \pi b_{\rm off} \bar{A} db_{\rm off}}{\displaystyle \int _0^{b_{\rm max}} 2 \pi b_{\rm off} db_{\rm off}}
  \mathrm{,}
  \label{equ:b_weight_def}
\end{equation}
which is called the $b_{\mathrm{off}}$-weighted average, where $b_{\mathrm{max}}$ is the maximum value of the impact parameter.
The maximum impact parameter is the sum of the characteristic radii, $(5/3)^{1/2} r_{\mathrm{gyr}}$, of the target and the projectile, where $r_{\mathrm{gyr}}$ is the gyration radius of the dust aggregate (Mukai et al. 1992; Wada et al. 2013).
The critical collisional fragmentation velocity, $v_{\rm fra}$, is defined as the collision velocity when the $b_{\mathrm{off}}$-weighted average mass of the largest remnant, $\langle M_{\rm lar} \rangle = \langle N_{\rm lar} \rangle m_{\rm mon}$, is equal to the initial mass of the target, $M_{\rm tar} = N_{\rm tar} m_{\rm mon}$.
Variables used in this paper are summarized in Table \ref{tab:variables}.

\begin{table}
  \centering
  \caption{Definition of Symbols Used in This Paper}
  \label{tab:variables}
  \begin{tabular}{ll}
    \hline
    \hline
    Symbol & Definition \\
    \hline
    $E_{\rm bre}$ & Energy for breaking a single contact between two dust monomers in the equilibrium position, $E_{\rm bre} = 6.1 \times 10^{-10} ~ \mathrm{erg}$ \\
    $v_{\rm bre}$ & Characteristic break-up velocity, $v_{\rm bre} \equiv (E_{\rm bre}/m_{\rm mon})^{1/2} = 3.8 ~ \mathrm{m ~ s^{-1}}$ \\
    $m_{\rm mon}$ & Mass of a dust monomer, $m_{\rm mon} = 4.2 \times 10^{-15} ~ \mathrm{g}$ \\
    $N$ & Number of dust monomers constituting a dust aggregate \\
    $M$ & Mass of a dust aggregate composed of dust monomers of $N$, $M = N m_{\rm mon}$ \\
    $N_{\rm tar}$ & Number of dust monomers constituting the target (Target is defined as the larger colliding dust aggregate) \\
    $M_{\rm tar}$ & Mass of the target, $M_{\rm tar} = N_{\rm tar} m_{\rm mon}$ \\
    $N_{\rm pro}$ & Number of dust monomers constituting the projectile (Projectile is defined as the smaller colliding dust aggregate) \\
    $M_{\rm pro}$ & Mass of the projectile, $M_{\rm pro} = N_{\rm pro} m_{\rm mon}$ \\
    $v_{\rm col}$ & Collision velocity between the target and the projectile \\
    $b_{\rm off}$ & Impact parameter \\
    $N_{\rm tot}$ & Sum of the monomer numbers constituting the target and the projectile, $N_{\rm tot} = N_{\rm tar} + N_{\rm pro}$ \\
    $M_{\rm tot}$ & Sum of the masses of the target and the projectile, $M_{\rm tot} = M_{\rm tar} + M_{\rm pro} = N_{\rm tot} m_{\rm mon}$ \\
    $v_{\rm fra}$ & Critical collisional fragmentation velocity defined as the lowest collision velocity for fragmentation \\
    $N_{\rm lar}$ & Number of dust monomers in the largest remnant \\
    $M_{\rm lar}$ & Mass of the largest remnant, $M_{\rm lar} = N_{\rm lar} m_{\rm mon}$ \\
    $N_{\rm 2nd}$ & Number of dust monomers in the second-largest remnant \\
    $M_{\rm 2nd}$ & Mass of the second-largest remnant, $M_{\rm 2nd} = N_{\rm 2nd} m_{\rm mon}$ \\
    $N_{\rm eje}$ & Total monomer number constituting ejecta except for the two large remnants, $N_{\rm eje} = N_{\rm tot} - (N_{\rm lar} + N_{\rm 2nd})$ \\
    $M_{\rm eje}$ & Total mass of ejecta except for the two large remnants, $M_{\rm eje} = M_{\rm tot} - (M_{\rm lar} + M_{\rm 2nd}) = N_{\rm eje} m_{\rm mon}$ \\
    $M_{\rm cum}$ & Cumulative mass distribution of fragments not larger than a given mass $M$, $M_{\rm cum}(M) = \int _{m_{\rm mon}}^M M n(M) dM$ \\
    { }           & (Cumulative number distribution of fragments not smaller than a given mass $M$ is given by $\int _M^{M_{\rm tot}} n(M) dM$) \\
    $n(M)$ & Integrand in the cumulative mass distribution of fragments, normalized by the fragment mass, $n(M) = (dM_{\rm cum} / dM) / M$ \\
    { }    & ($n(M) dM$ is the number of fragments with a mass between $M$ and $M + dM$) \\
    $A_{\rm cum}$ & Coefficient of the fitting formulae (\ref{equ:fitting_NcumoNtot}) and (\ref{equ:fitting_NcumoNtot_model}) \\
    $B_{\rm cum}$ & Power-law index of the fitting formulae (\ref{equ:fitting_NcumoNtot}) and (\ref{equ:fitting_NcumoNtot_model}) \\
    $C_{\rm cum}$ & Mass fraction of fragments in the individual-monomer state, normalized by the total mass of colliding bodies, and \\
    { }           & intercept of the fitting formulae (\ref{equ:fitting_NcumoNtot}) and (\ref{equ:fitting_NcumoNtot_model}) \\
    $N_{\rm mon}$ & Number of fragments in the individual-monomer state, $N_{\rm mon} = C_{\rm cum} N_{\rm tot}$ \\
    $M_{\rm mon}$ & Total mass of fragments in the individual-monomer state, $M_{\rm mon} = C_{\rm cum} M_{\rm tot} = N_{\rm mon} m_{\rm mon}$ \\
    $M_{\rm pow}$ & Upper limit of the fragment mass with $\langle M_{\rm cum} \rangle (M) \le \langle M_{\rm eje} \rangle$ \\
    $Q_{\rm Ctot}$ & Power-law index for the dependence of $C_{\rm cum}$ on $N_{\rm tot}$ (Equations (\ref{equ:fitting_Ccum}) and (\ref{equ:fitting_QCtot})) \\
    $S(N)$        & Geometric cross section of a dust grain with $M = N m_{\rm mon}$, with assuming that all dust aggregates are \\
    { }           & spherical-shaped compact dust grains for simplicity, $S(N) = \pi (0.1 N^{1/3} ~ \mathrm{\micron})^2$ \\
    $S_{\rm cum}$ & Ratio of the cumulative geometric cross section of the fragments not larger than a given mass $M$ to the geometric \\
    { }           & cross section of a dust aggregate with $M = M_{\rm tot}$, $S_{\rm cum}(M) = \int _{m_{\rm mon}}^M S(N) n(M) dM / S(N_{\rm tot}) = \int _{m_{\rm mon}}^M (M / M_{\rm tot})^{2/3} n(M) dM$ \\
    \hline
  \end{tabular}
\end{table}

In this study, 16,160 runs were newly carried out in addition to those in Paper I, and numerical results of total 24,320 runs were used for the analysis (Table \ref{tab:list}).

\begin{table}
  \centering
  \caption{List of Sampling Points in Input Parameters ($N_{\rm tar}$, $N_{\rm pro}$, $v_{\rm col}$, and $b_{\rm off}$)}
  \label{tab:list}
  \begin{tabular}{cccc}
    \hline
    \hline
    \multicolumn{4}{c}{Simulations with the surface energy of $100 ~ \mathrm{mJ ~ m^{-2}}$} \\
    \hline
    $N_{\rm tar}$ & $N_{\rm pro}$ & $v_{\rm col}$ [$\mathrm{m ~ s^{-1}}$] & $b_{\rm off} / b_{\mathrm{max}}$ \\
    \hline
    65536  &  65536,  52428,  43690,  37449, & 15,  22,  31,  44,  62, & 0.05, 0.15, $\ldots$, 0.95 \\
    { }    &  32768,  26214,  21845,  16384, & 87, 123, 174, 246, 348  & { } \\
    { }    &  10922                          & { }                     & { } \\
    \hline
    131072 & 131072, 104857,  87381,  74898, & 15,  22,  31,  44,  62, & 0.05, 0.15, $\ldots$, 0.95 \\
    { }    &  65536,  52428,  43690,  32768, & 87, 123, 174, 246, 348  & { } \\
    { }    &  21845,  16384,  10922,   8192, & { }                     & { } \\
    { }    &   5461,   4096,   2730,   2048  & { }                     & { } \\
    \hline
    262144 & 262144, 209715, 174762, 149796, & 15,  22,  31,  44,  62, & 0.05, 0.15, $\ldots$, 0.95 \\
    { }    & 131072, 104857,  87381,  65536, & 87, 123, 174, 246, 348  & { } \\
    { }    &  43690,  32768,  21845,  16384, & { }                     & { } \\
    { }    &  10922,   8192,   5461,   4096  & { }                     & { } \\
    \hline
    524288 &          65536,  43690,  32768, & 15,  22,  31,  44,  62, & 0.05, 0.15, $\ldots$, 0.95 \\
    { }    &  21845,  16384,  10922,   8192  & 87, 123, 174, 246, 348  & { } \\
    \hline
    \multicolumn{4}{c}{Simulations with the surface energy of $50 ~ \mathrm{mJ ~ m^{-2}}$ (Section \ref{sec:dis_mat_ice})} \\
    \hline
    65536  &  65536,  43690,  32768,  21845, & 8.6, 12, 17, 24, & 0.05, 0.15, $\ldots$, 0.95 \\
    { }    &  16384                          & 35,  49, 69, 98  & { } \\
    \hline
    131072 & 131072,  87381,  65536,  43690, & 8.6, 12, 17, 24, & 0.05, 0.15, $\ldots$, 0.95 \\
    { }    &  32768,  16384,   8192,   4096  & 35,  49, 69, 98  & { } \\
    \hline
    262144 & 262144, 174762, 131072          & 8.6, 12, 17, 24, & 0.05, 0.15, $\ldots$, 0.95 \\
    { }    & { }                             & 35,  49, 69, 98  & { } \\
    \hline
  \end{tabular}
\end{table}

Simulations with higher mass ratios and/or higher collision velocities require a larger amount of particles in order to achieve the numerical convergence.
In this paper, results of simulations with $N_{\rm tar} =$ 131072 and 262144 are employed in fits for $M_{\rm tar} / M_{\rm pro}$ ($= (N_{\rm tar} m_{\rm mon}) / (N_{\rm pro} m_{\rm mon})$) $= N_{\rm tar} / N_{\rm pro} \le 6$, with $N_{\rm tar} =$ 262144 and 524288 for $8 \le N_{\rm tar} / N_{\rm pro} \le 24$, and with $N_{\rm tar} = 524288$ for $N_{\rm tar} / N_{\rm pro} \ge 32$, except for fits of mass distributions of tiny fragments (Section \ref{sec:res_small}).
The fragment model presented in this paper should be used within the parameter ranges of our numerical simulations, i.e., $v_{\rm col} \le 3.5 \times 10^2 ~ \mathrm{m ~ s^{-1}}$ and $N_{\rm tar} / N_{\rm pro} \le 64$.

\section{Outline of Our Fragment Model} \label{sec:mod}

\begin{figure}
  \gridline{\fig{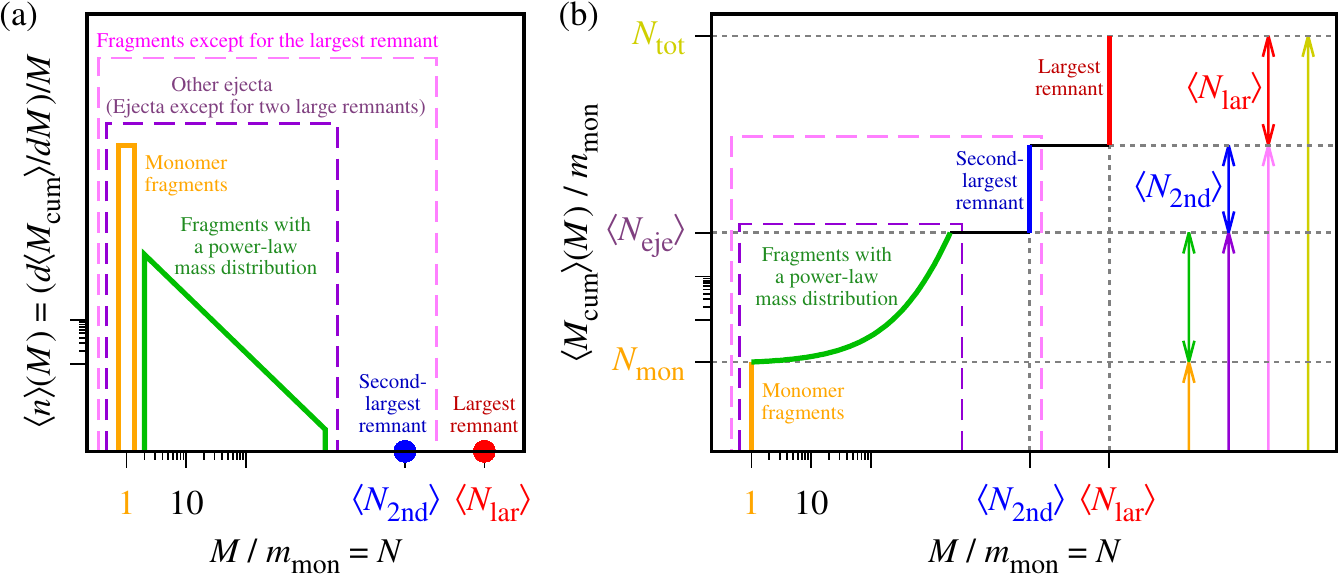}{0.8\textwidth}{}}
  \gridline{\fig{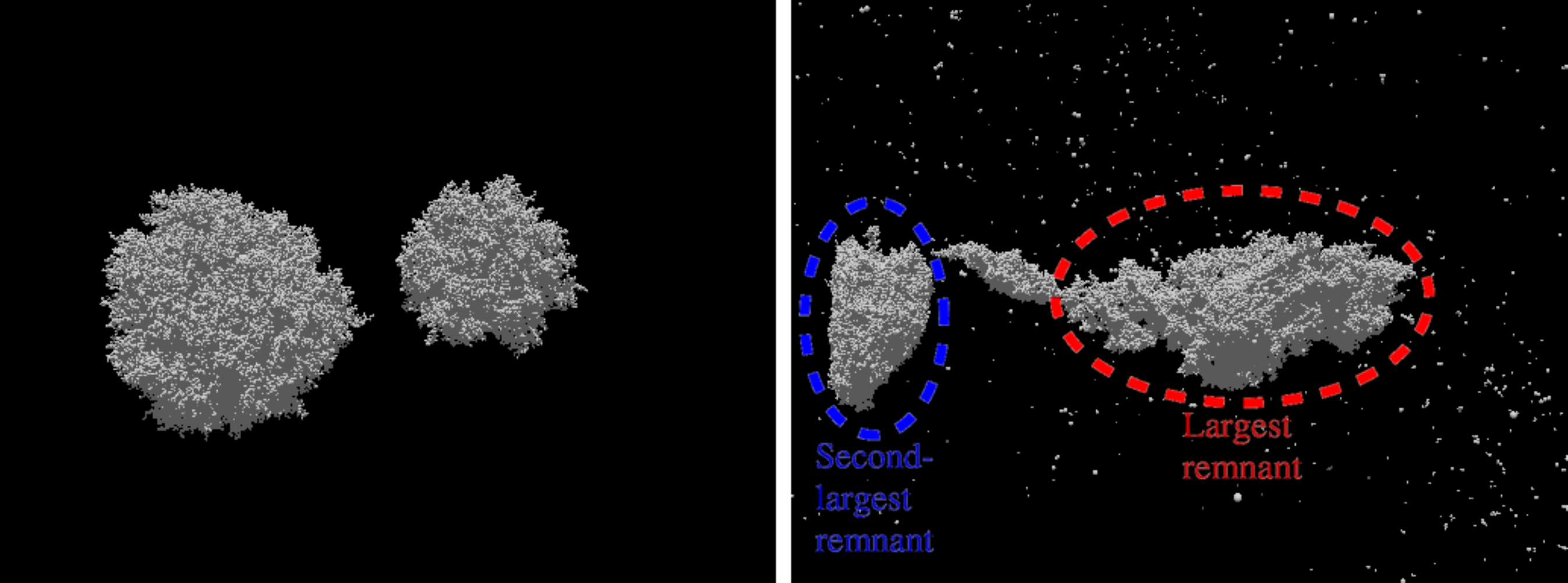}{0.8\textwidth}{}}
  \caption{Schematic pictures of the mass distribution of the fragments (upper panels), and snapshots of an offset collision with $b_{\rm off} / b_{\mathrm{max}} = 0.25$ between a target of $N_{\rm tar} = 65536$ and a projectile of $N_{\rm pro} = 21845$ (i.e., $N_{\rm tar} / N_{\rm pro} = 3$) for $v_{\rm col} = 87 ~\mathrm{m ~ s^{-1}}$ (lower panels), respectively. (Upper panel (a)) Mass distribution of the fragments, $\langle n \rangle (M)$ (solid lines and circles). The red circle and the blue circle indicate the normalized masses of the largest remnant and the second-largest remnant, $\langle M_{\rm lar} \rangle / m_{\rm mon} = \langle N_{\rm lar} \rangle$ and $\langle M_{\rm 2nd} \rangle / m_{\rm mon} = \langle N_{\rm 2nd} \rangle$, respectively. The magenta dashed line and the purple dashed line represent ranges of the fragments except for the largest remnant and of the ejecta except for the two large remnants, respectively. The green solid line and the orange solid line represent fragments with the power-law mass distribution and individual dust monomers, respectively. (Upper panel (b)) Cumulative mass distribution of fragments not larger than a given mass $M$, normalized by the monomer mass, $\langle M_{\rm cum} \rangle (M) / m_{\rm mon}$, against the normalized fragment mass $M / m_{\rm mon} = N$ (solid line). Colors are the same as in panel (a). (Lower panels) Initial condition (left) and collisional outcome (right). The largest fragment surrounded by the red dotted line is the largest remnant drawn in red in the upper panels. The largest fragment surrounded by the blue dotted line is the second-largest remnant drawn in blue in the upper panels. Other fragments except for those two large remnants are other ejecta drawn in purple in the upper panels.}
  \label{fig:diagram}
\end{figure}

The purpose of this paper is to model the mass distribution of the fragments for the collision velocity and the mass ratio between colliding dust aggregates by numerical fitting.
As shown in Figure \ref{fig:diagram}, we divide all the fragments into three classes, which are the largest remnant (drawn in red in Figure \ref{fig:diagram}), the second-largest remnant (hereafter, called the second remnant; blue in Figure \ref{fig:diagram}), and other ejecta except for these two large remnants (hereafter, called the other ejecta; purple in Figure \ref{fig:diagram}).
In addition, the other ejecta (purple) are classified into fragments with the power-law mass distribution (green in Figure \ref{fig:diagram}) and fragments to be individual dust monomers (orange in Figure \ref{fig:diagram}) in Section \ref{sec:res_small}.
For modeling of the mass distribution of the fragments, we need to derive analytical expressions of the mass of the largest remnant, $\langle M_{\rm lar} \rangle = \langle N_{\rm lar} \rangle m_{\rm mon}$ (red), the mass of the second remnant, $\langle M_{\rm 2nd} \rangle = \langle N_{\rm 2nd} \rangle m_{\rm mon}$ (blue), and the cumulative mass distribution of all fragments, $\langle M_{\rm cum} \rangle (M)$ (a whole solid line in Figure \ref{fig:diagram}(b)), for the collision velocity and the mass ratio between the two colliding bodies, $v_{\rm col}$ and $M_{\rm tar} / M_{\rm pro} = N_{\rm tar} / N_{\rm pro}$.
We determined the second remnant mass $\langle M_{\rm 2nd} \rangle$ (blue) and the total mass of the other ejecta, $\langle M_{\rm eje} \rangle \equiv M_{\rm tot} - (\langle M_{\rm lar} \rangle + \langle M_{\rm 2nd} \rangle) = \langle N_{\rm eje} \rangle m_{\rm mon}$ (purple), by direct numerical fitting, and derive the largest remnant mass $\langle M_{\rm lar} \rangle$ (red) from
\begin{equation}
  \langle M_{\rm lar} \rangle = M_{\rm tot} - (\langle M_{\rm 2nd} \rangle + \langle M_{\rm eje} \rangle )
  \mathrm{,}
  \label{equ:M_lar}
\end{equation}
where $M_{\rm tot} = N_{\rm tot} m_{\rm mon}$ is the total mass of the colliding bodies (yellow in Figure \ref{fig:diagram}(b)), and $M_{\rm tot} - \langle M_{\rm lar} \rangle = (N_{\rm tot} - \langle N_{\rm lar} \rangle) m_{\rm mon}$ is drawn in magenta in Figure \ref{fig:diagram}.
This procedure can assure $\langle M_{\rm lar} \rangle$ (red) $\ge$ $\langle M_{\rm 2nd} \rangle$ (blue), whereas additional technical conditions will also be imposed (Section \ref{sec:res_large}).

Readers may notice that alternative methods are possible to determine the three components of the fragments: (i) The second remnant mass $\langle M_{\rm 2nd} \rangle$ (blue) is derived from the numerically fitted largest remnant mass $\langle M_{\rm lar} \rangle$ (red) and total other ejecta mass $\langle M_{\rm eje} \rangle$ (purple).
(ii) The total other ejecta mass $\langle M_{\rm eje} \rangle$ (purple) is derived from the numerically fitted largest remnant mass $\langle M_{\rm lar} \rangle$ (red) and second remnant mass $\langle M_{\rm 2nd} \rangle$ (blue).
We also tested these alternative procedures and found that our current one is better than them.
For example, when taking the method (i), we found some cases that yield very large $\langle M_{\rm 2nd} \rangle$ (blue) $\rightarrow$ $\infty$ for $v_{\rm col} \rightarrow \infty$, although physically $\langle M_{\rm 2nd} \rangle$ (blue) should be $\rightarrow m_{\rm mon}$ for $v_{\rm col} \rightarrow \infty$.
In method (ii), some cases give negative $\langle M_{\rm eje} \rangle$ (purple) locally for low to middle $v_{\rm col}$, because this procedure does not impose to conserve the total mass of all fragments.

\section{Results} \label{sec:res}

In this section, we present analytic formulae that describe the physical properties of the $N$-body simulations.
How to use our fragment model is summarized in Table \ref{tab:procedure} at the end of this section.

\subsection{Masses of Large Remnants and Other Ejecta} \label{sec:res_large}

\begin{figure}
  \plotone{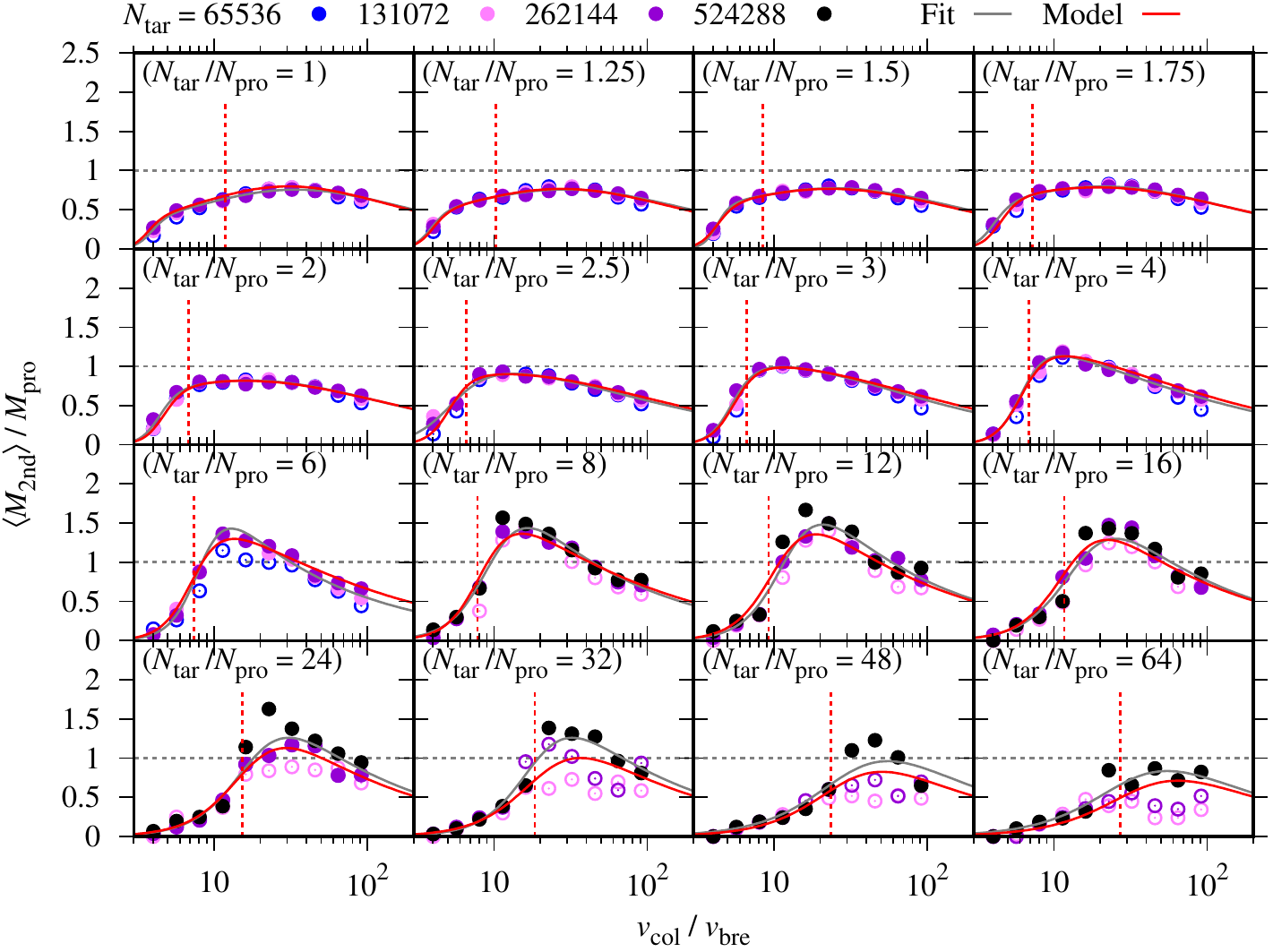}
  \caption{The $b_{\mathrm{off}}$-weighted average mass of the second-largest remnant, normalized by the projectile mass, $\langle M_{\rm 2nd} \rangle / M_{\rm pro} = \langle N_{\rm 2nd} \rangle / N_{\rm pro}$, against the collision velocity normalized by the characteristic break-up velocity, $v_{\rm col} / v_{\rm bre}$ (circles). Filled circles indicate data points employed in fits, while open circles indicate those not employed. Colors except for gray and red represent the monomer number of the target, $N_{\rm tar}$. Gray solid lines mark fitting relations derived from Equation (\ref{equ:fitting_N2ndoNpro}) with assuming constant $B_{\rm SH} = -0.529$. Red solid lines indicate analytic formulae derived from Equations (\ref{equ:fitting_ASLandBSL_model}) and (\ref{equ:fitting_N2ndoNpro_model}). Horizontal gray dotted lines denote $\langle M_{\rm 2nd} \rangle = M_{\rm pro}$. Vertical red dotted lines represent $v_{\rm col} = v_{\rm fra}$ obtained from our fragment model. Different panels correspond to results of different initial mass ratios $M_{\rm tar} / M_{\rm pro} = N_{\rm tar} / N_{\rm pro}$.}
  \label{fig:fitting_vcob_N2ndoNpro_Ntar_Ntop_ylin_fit_model}
\end{figure}

Figure \ref{fig:fitting_vcob_N2ndoNpro_Ntar_Ntop_ylin_fit_model} shows the $b_{\mathrm{off}}$-weighted average mass of the second-largest remnant, normalized by the projectile mass, $\langle M_{\rm 2nd} \rangle / M_{\rm pro} = \langle N_{\rm 2nd} \rangle / N_{\rm pro}$, against the collision velocity normalized by the characteristic break-up velocity, $v_{\rm col} / v_{\rm bre}$, for different initial mass ratios $M_{\rm tar} / M_{\rm pro}$ ($= N_{\rm tar} / N_{\rm pro}$).
Hereafter, the mass ratio is written as $N_{\rm tar} / N_{\rm pro}$ since $N_{\rm tar}$ and $N_{\rm pro}$ are input parameters of our simulations.
In any mass ratio, $\langle M_{\rm 2nd} \rangle \ll M_{\rm tot}$ for $v_{\rm col} \ll v_{\rm fra}$, because the collisional sticking of dust aggregates produces a single large body and small $\langle M_{\rm 2nd} \rangle$.
Meanwhile high velocity impacts with $v_{\rm col} \gg v_{\rm fra}$ also give small $\langle M_{\rm 2nd} \rangle$ due to collisional fragmentation.
For $N_{\rm tar} / N_{\rm pro} \le 2$, the mass of the second remnant shows a nearly flat dependence on $v_{\rm col}$ with $\langle M_{\rm 2nd} \rangle / M_{\rm pro} < 1$ around the critical fragmentation velocity, while there is a specific peak with $\langle M_{\rm 2nd} \rangle > M_{\rm pro}$ for $N_{\rm tar} / N_{\rm pro} \gtrsim 6$.
The results for $2.5 \le N_{\rm tar} / N_{\rm pro} \lesssim 4$ have the maximum values of $\langle M_{\rm 2nd} \rangle \approx M_{\rm pro}$, which corresponds to the transition from the flat dependence to the profile with a peak.
Paper I showed that the second remnant is larger than the projectile due to the mass transfer from the target to the projectile for $N_{\rm tar} / N_{\rm pro} \gtrsim 3$.
The peak in $\langle M_{\rm 2nd} \rangle / M_{\rm pro}$ is a characteristic feature of the mass transfer.

Figure \ref{fig:fitting_vcob_N2ndoNpro_Ntar_Ntop_ylin_fit_model} shows that the second remnant mass $\langle M_{\rm 2nd} \rangle$ consists of two or three velocity components.
Taking into account the velocity dependence explained above, we assume the fitting formulae of the normalized second remnant mass as
\begin{eqnarray}
  \frac{\langle M_{\rm 2nd} \rangle}{M_{\rm pro}} = \left \{ \begin{array}{ll}
    \left \{ \left [ A_{\rm SL} \left ( \displaystyle \frac{v_{\rm col}}{v_{\rm bre}} \right ) ^{B_{\rm SL}} \right ] ^{-1}          + \left [ A_{\rm SM} \left ( \displaystyle \frac{v_{\rm col}}{v_{\rm bre}} \right ) ^{B_{\rm SM}} \right ] ^{-1}  + \left [ A_{\rm SH} \left ( \displaystyle \frac{v_{\rm col}}{v_{\rm bre}} \right ) ^{B_{\rm SH}} \right ] ^{-1} \right \} ^{-1} & \mathrm{for} ~ \mathrm{the ~ low ~ mass ~ ratio} \\
    \left \{ \left [ A_{\rm SL} \left ( \displaystyle \frac{v_{\rm col}}{v_{\rm bre}} \right ) ^{B_{\rm SL}} \right ] ^{-1} \phantom{+ \left [ A_{\rm SM} \left ( \displaystyle \frac{v_{\rm col}}{v_{\rm bre}} \right ) ^{B_{\rm SM}} \right ] ^{-1}} + \left [ A_{\rm SH} \left ( \displaystyle \frac{v_{\rm col}}{v_{\rm bre}} \right ) ^{B_{\rm SH}} \right ] ^{-1} \right \} ^{-1} & \mathrm{for} ~ \mathrm{the ~ high ~ mass ~ ratio} \\
  \end{array} \right .
  \mathrm{,}
  \label{equ:fitting_N2ndoNpro}
\end{eqnarray}
where $A_{\rm SL}$, $A_{\rm SM}$, $A_{\rm SH}$, $B_{\rm SL}$, $B_{\rm SM}$, and $B_{\rm SH}$ are the fitting parameters given by functions of the mass ratio $N_{\rm tar} / N_{\rm pro}$, and the suffixes L, M, and H indicate low-, middle-, and high-velocity components, respectively.
As shown in Appendix \ref{sec:app_spe_pro}, a constant power-law index $B_{\rm SH} = -0.529$ in the high-velocity regime nicely explains the simulation results irrespective of the mass ratio.
Re-fitting the numerical results by Equation (\ref{equ:fitting_N2ndoNpro}) with the obtained $B_{\rm SH}$, we confirm that the second remnant mass $\langle M_{\rm 2nd} \rangle$ contains the shallow power-law function given by $A_{\rm SM}$ and $B_{\rm SM}$ in the middle-velocity range for $N_{\rm tar} / N_{\rm pro} \le 4$.
In contrast, we also find that, for $N_{\rm tar} / N_{\rm pro} \ge 6$, the contribution from this middle velocity component is negligible.

\begin{figure}
  \plotone{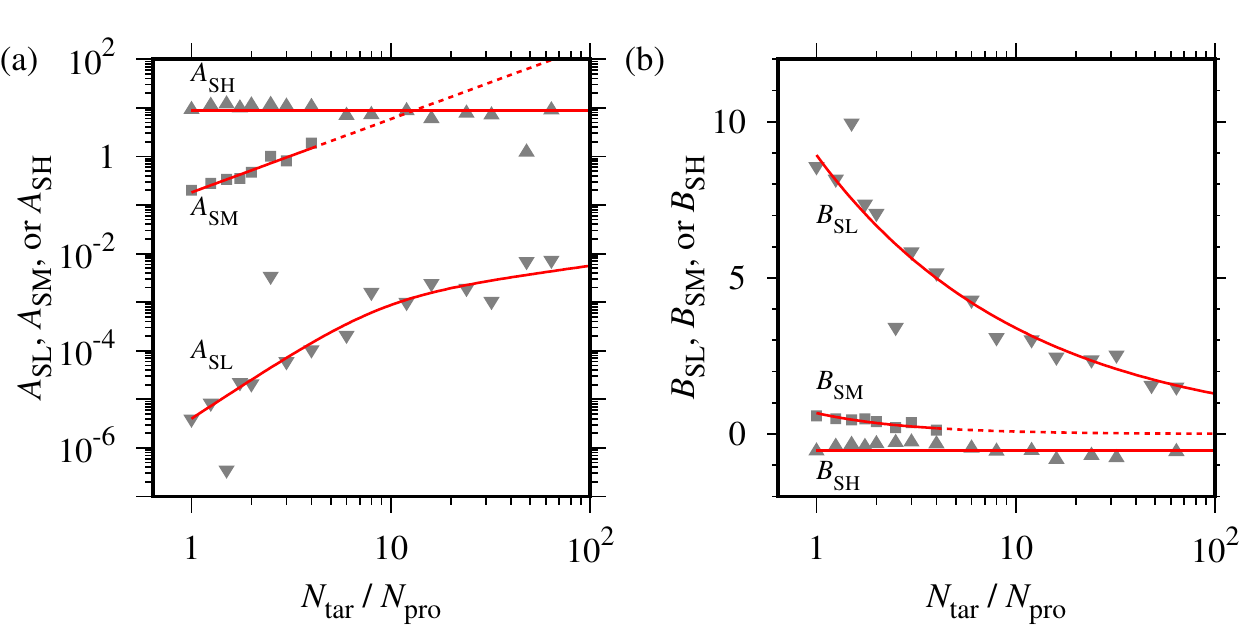}
  \caption{Coefficients (a) and power-law indices (b) of each term of the fitting formulae (\ref{equ:fitting_N2ndoNpro}) for re-fits with assuming constant $B_{\rm SH} = -0.529$ as functions of the initial mass ratio, $N_{\rm tar} / N_{\rm pro}$. Down-pointing triangles, squares, and up-pointing triangles indicate low-, middle-, and high-velocity components, respectively. Red lines mark analytic formulae given by fitting relations of fitting parameters, derived from Equation (\ref{equ:fitting_ASLandBSL_model}).}
  \label{fig:fitting_Ntop_fittingparameters_N2nd_model}
\end{figure}

Figure \ref{fig:fitting_Ntop_fittingparameters_N2nd_model} shows coefficients and power-law indices of Equation (\ref{equ:fitting_N2ndoNpro}) for the obtained constant $B_{\rm SH}$ as functions of the initial mass ratio, which are explicitly expressed as
\begin{eqnarray}
  \begin{array}{l}
    A_{\rm SL} = \left \{ \left [ 4.04 \times 10^{-6} \left ( \displaystyle \frac{N_{\rm tar}}{N_{\rm pro}} \right ) ^{2.70} \right ] ^{-1} + \left [ 4.13 \times 10^{-4} \left ( \displaystyle \frac{N_{\rm tar}}{N_{\rm pro}} \right ) ^{0.568} \right ] ^{-1} \right \} ^{-1} \mathrm{,} \\
    B_{\rm SL} =                  8.93                \left ( \displaystyle \frac{N_{\rm tar}}{N_{\rm pro}} \right ) ^{-0.420} \mathrm{,} \\
    A_{\rm SM} = 0.181 \left ( \displaystyle \frac{N_{\rm tar}}{N_{\rm pro}} \right ) ^{1.51} \mathrm{,} \\
    B_{\rm SM} = 0.673 \left ( \displaystyle \frac{N_{\rm tar}}{N_{\rm pro}} \right ) ^{-0.926} \mathrm{,} \\
    A_{\rm SH} = 8.69 \mathrm{,} \\
    \mathrm{and} \\
    B_{\rm SH} = -0.529 \mathrm{.}
  \end{array}
  \label{equ:fitting_ASLandBSL_model}
\end{eqnarray}
Equation (\ref{equ:fitting_ASLandBSL_model}) confirms that the contribution of the second term in Equation (\ref{equ:fitting_N2ndoNpro}) is almost negligible for collisions with high-mass ratio $N_{\rm tar} / N_{\rm pro} \ge 6$.
Thus, we artificially modify the fitting formulae of the normalized second remnant mass (Equation (\ref{equ:fitting_N2ndoNpro})) by
\begin{equation}
  \frac{\langle M_{\rm 2nd} \rangle}{M_{\rm pro}} = \left \{ \left [ A_{\rm SL} \left ( \frac{v_{\rm col}}{v_{\rm bre}} \right ) ^{B_{\rm SL}} \right ] ^{-1} + \left [ A_{\rm SM} \left ( \frac{v_{\rm col}}{v_{\rm bre}} \right ) ^{B_{\rm SM}} \right ] ^{-1} + \left [ A_{\rm SH} \left ( \frac{v_{\rm col}}{v_{\rm bre}} \right ) ^{B_{\rm SH}} \right ] ^{-1} \right \} ^{-1}
  \mathrm{.}
  \label{equ:fitting_N2ndoNpro_model}
\end{equation}
This modified fitting formula, however, does not ensure $\langle M_{\rm 2nd} \rangle \ge m_{\rm mon}$ in an extremely high-velocity collision, although we assume that all dust aggregates are constructed from spherical dust monomers with masses of $m_{\rm mon}$.
In order to ensure $\langle M_{\rm 2nd} \rangle \ge m_{\rm mon}$ and $\langle M_{\rm 2nd} \rangle \rightarrow m_{\rm mon}$ for $v_{\rm col} \rightarrow \infty$, the high-velocity component of Equation (\ref{equ:fitting_N2ndoNpro_model}), $A_{\rm SH} (v_{\rm col} / v_{\rm bre})^{B_{\rm SH}}$, should be replaced with $A_{\rm SH} (v_{\rm col} / v_{\rm bre})^{B_{\rm SH}} + (1 / N_{\rm pro})$.
This intercept $1 / N_{\rm pro}$ contributes to the second remnant mass only for the extremely high collision velocity, $v_{\rm col} \gg 10^2 ~ \mathrm{m ~ s^{-1}}$.
However, in such an extremely high-velocity collision, the shape of dust monomers will get warped and will not satisfy the assumption of the spherical dust monomers (Tanaka et al. 2012; Mill\'{a}n et al. 2016).
Thus, we can neglect the intercept of the high-velocity component, $1 / N_{\rm pro}$, for the collision velocity we are considering, $v_{\rm col} \le 3.5 \times 10^2 ~ \mathrm{m ~ s^{-1}}$, as long as $1 \ll N_{\rm pro} \rightarrow \infty$.
The fitting results of Equations (\ref{equ:fitting_ASLandBSL_model}) and (\ref{equ:fitting_N2ndoNpro_model}) are shown in Figure \ref{fig:fitting_vcob_N2ndoNpro_Ntar_Ntop_ylin_fit_model} (red lines), which nicely reproduce the overall trends of the numerical results (data points).

\begin{figure}
  \plotone{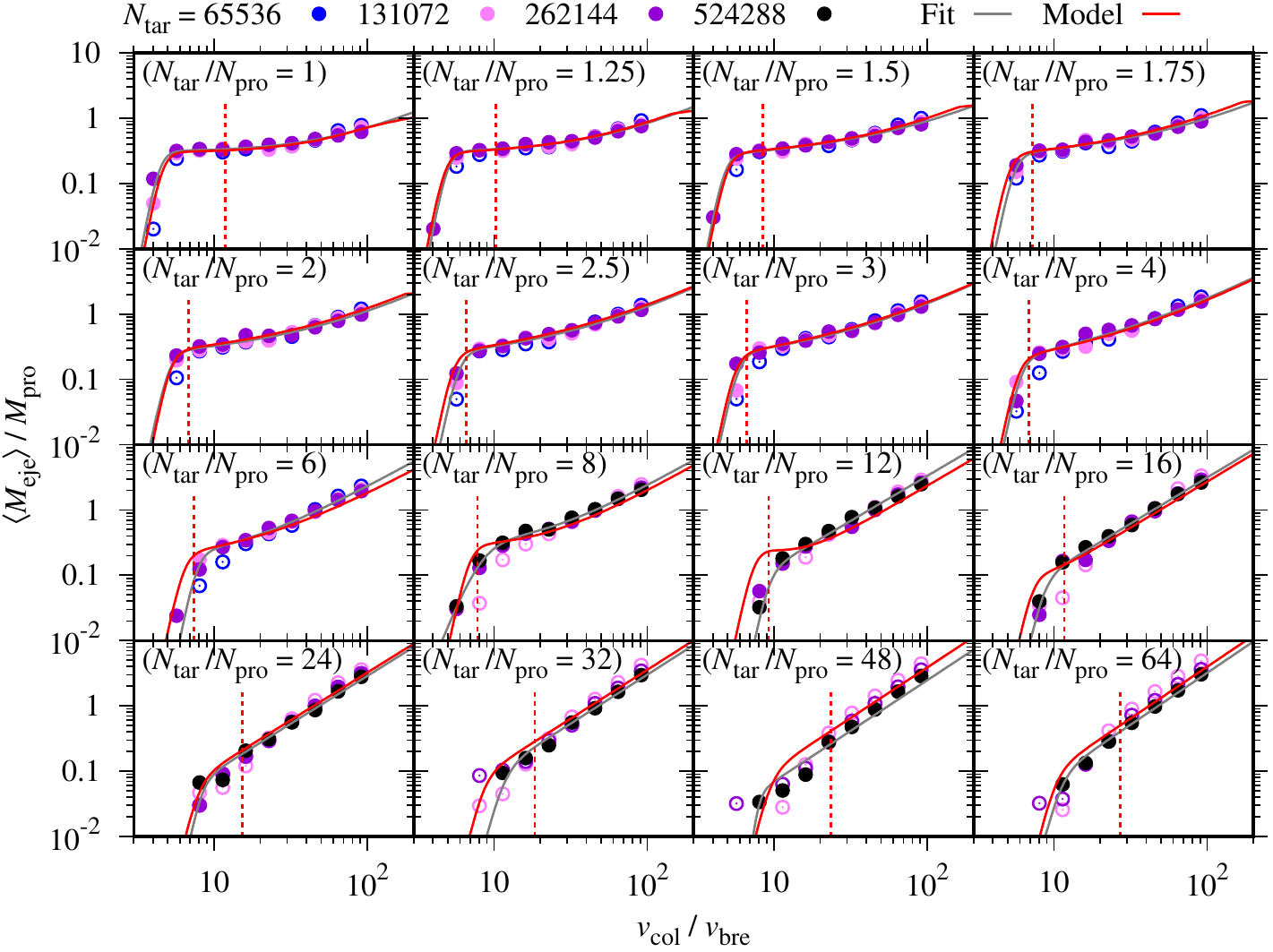}
  \caption{Average of the total mass weighted over $b_{\mathrm{off}}$ of ejecta except for the two large remnants, normalized by the projectile mass, $\langle M_{\rm eje} \rangle / M_{\rm pro} = [M_{\rm tot} - (\langle M_{\rm lar} \rangle + \langle M_{\rm 2nd} \rangle)] / M_{\rm pro} = \langle N_{\rm eje} \rangle / N_{\rm pro}$, against the normalized collision velocity, $v_{\rm col} / v_{\rm bre}$ (circles). Filled and open circles and colors except for gray and red are the same as Figure \ref{fig:fitting_vcob_N2ndoNpro_Ntar_Ntop_ylin_fit_model}. Gray solid lines mark fitting relations derived from Equation (\ref{equ:fitting_E1a2oNpro}), with assuming the power-law index as the simple power law, $B_{\rm EH} = 1.12 (N_{\rm tar} / N_{\rm pro})^{0.0834}$. Red solid lines indicate analytic formulae derived from Equations (\ref{equ:fitting_AELandBEL_model}) and (\ref{equ:fitting_E1a2oNpro_model}). Vertical red dotted lines denote $v_{\rm col} = v_{\rm fra}$ obtained from our fragment model. Different panels correspond to results of different initial mass ratios.}
  \label{fig:fitting_vcob_E1a2oNpro_Ntar_Ntop_ylog_fit_model}
\end{figure}

Figure \ref{fig:fitting_vcob_E1a2oNpro_Ntar_Ntop_ylog_fit_model} shows the $b_{\mathrm{off}}$-weighted average total mass of other ejecta, normalized by the projectile mass, $[M_{\rm tot} - (\langle M_{\rm lar} \rangle + \langle M_{\rm 2nd} \rangle)] / M_{\rm pro} \equiv \langle M_{\rm eje} \rangle / M_{\rm pro} = \langle N_{\rm eje} \rangle / N_{\rm pro}$, against the normalized collision velocity, $v_{\rm col} / v_{\rm bre}$, for different initial mass ratios, $N_{\rm tar} / N_{\rm pro}$.
Figure \ref{fig:fitting_vcob_E1a2oNpro_Ntar_Ntop_ylog_fit_model} shows that for collisions with low-mass ratio $N_{\rm tar} / N_{\rm pro} \le 8$ the fraction of the small fragments does not increase rapidly in the middle-velocity range.
In contrast, for high-mass ratio $N_{\rm tar} / N_{\rm pro} \ge 12$ this middle-velocity component with slow dependence almost disappears.
The mass ratio when the transition occurs is slightly different from that for $\langle M_{\rm 2nd} \rangle$ (Figure \ref{fig:fitting_vcob_N2ndoNpro_Ntar_Ntop_ylin_fit_model}).
Figure \ref{fig:fitting_vcob_E1a2oNpro_Ntar_Ntop_ylog_fit_model} also indicates that collisions with mass transfer give small $\langle M_{\rm eje} \rangle / M_{\rm pro}$; if we evaluate $\langle M_{\rm eje} \rangle$ to the total mass of the two colliding objects, we obtain $\langle M_{\rm eje} \rangle = 0.089 M_{\rm tot}$ for $N_{\rm tar} = 262144$, $N_{\rm tar} / N_{\rm pro} = 3$ and $v_{\rm col} = 44 ~\mathrm{m ~ s^{-1}}$ for example. Figures \ref{fig:fitting_vcob_N2ndoNpro_Ntar_Ntop_ylin_fit_model} and \ref{fig:fitting_vcob_E1a2oNpro_Ntar_Ntop_ylog_fit_model} show that collisions with mass transfer, i.e., collisions with $\langle M_{\rm 2nd} \rangle > M_{\rm pro}$, yield $\langle M_{\rm eje} \rangle \lesssim 0.1 M_{\rm tot} = 0.1 (1 + N_{\rm tar} / N_{\rm pro}) M_{\rm pro}$.

Figure \ref{fig:fitting_vcob_E1a2oNpro_Ntar_Ntop_ylog_fit_model} shows that the total other ejecta mass $\langle M_{\rm eje} \rangle$ also consists of multiple velocity components, similarly to $\langle M_{\rm 2nd} \rangle$; we adopt the following fitting formulae:
\begin{eqnarray}
  \frac{\langle M_{\rm eje} \rangle}{M_{\rm pro}} = \left \{ \begin{array}{ll}
    \left \{ \left [ A_{\rm EL} \left ( \displaystyle \frac{v_{\rm col}}{v_{\rm bre}} \right ) ^{B_{\rm EL}} \right ] ^{-1} + \left [          A_{\rm EM} \left ( \displaystyle \frac{v_{\rm col}}{v_{\rm bre}} \right ) ^{B_{\rm EM}} +    A_{\rm EH} \left ( \displaystyle \frac{v_{\rm col}}{v_{\rm bre}} \right ) ^{B_{\rm EH}} \right ] ^{-1} \right \} ^{-1} & \mathrm{for} ~ \mathrm{the ~ low ~ mass ~ ratio} \\
    \left \{ \left [ A_{\rm EL} \left ( \displaystyle \frac{v_{\rm col}}{v_{\rm bre}} \right ) ^{B_{\rm EL}} \right ] ^{-1} + \left [ \phantom{A_{\rm EM} \left ( \displaystyle \frac{v_{\rm col}}{v_{\rm bre}} \right ) ^{B_{\rm EM}} +} ~ A_{\rm EH} \left ( \displaystyle \frac{v_{\rm col}}{v_{\rm bre}} \right ) ^{B_{\rm EH}} \right ] ^{-1} \right \} ^{-1} & \mathrm{for} ~ \mathrm{the ~ high ~ mass ~ ratio} \\
  \end{array} \right .
  \mathrm{,}
  \label{equ:fitting_E1a2oNpro}
\end{eqnarray}
where $A_{\rm EL}$, $A_{\rm EM}$, $A_{\rm EH}$, $B_{\rm EL}$, $B_{\rm EM}$, and $B_{\rm EH}$ are the fitting parameters that depend on the mass ratio $N_{\rm tar} / N_{\rm pro}$.
As shown in Appendix \ref{sec:app_spe_pro}, we firstly derive the power-law index in the high-velocity regime, $B_{\rm EH} = 1.12 (N_{\rm tar} / N_{\rm pro})^{0.0834}$, from numerical fitting.
Using the obtained $B_{\rm EH}$, we further derive the other parameters by numerical fitting.
We find that the total other ejecta mass $\langle M_{\rm eje} \rangle$ contains the middle-velocity component with a shallow power-law function for $N_{\rm tar} / N_{\rm pro} \le 8$ while it is not important for $N_{\rm tar} / N_{\rm pro} \ge 12$.

\begin{figure}
  \plotone{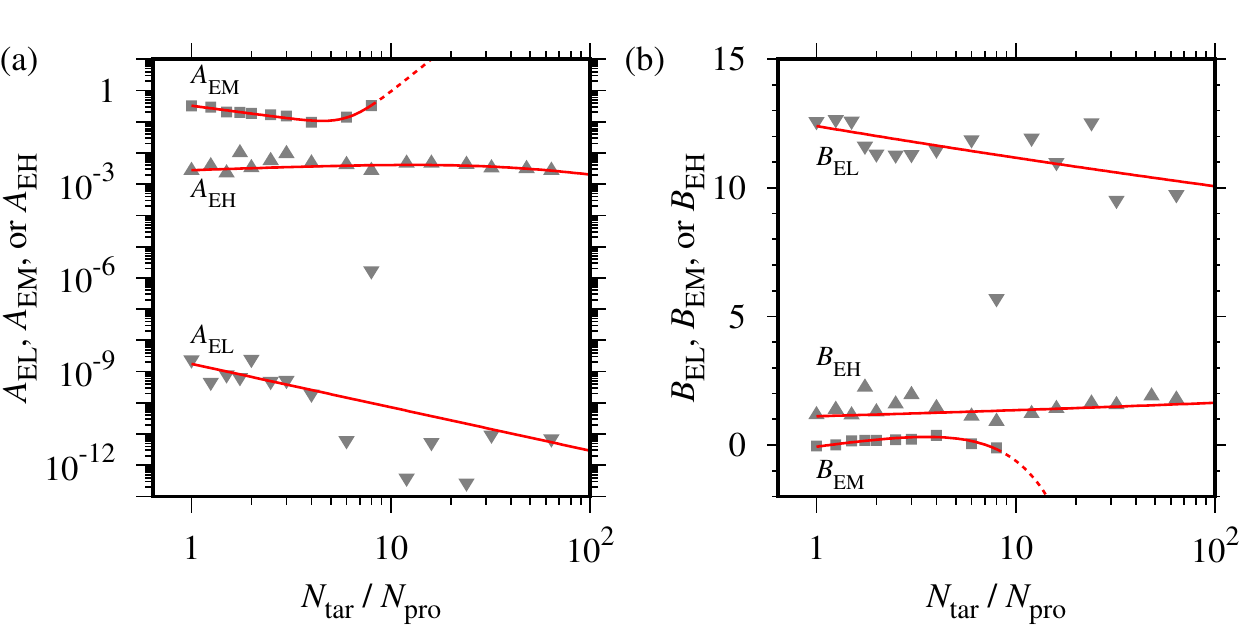}
  \caption{Coefficients (a) and power-law indices (b) of each term of the fitting formulae (\ref{equ:fitting_E1a2oNpro}) for re-fits with assuming $B_{\rm EH} = 1.12 (N_{\rm tar} / N_{\rm pro})^{0.0834}$ as functions of the initial mass ratio, $N_{\rm tar} / N_{\rm pro}$. Symbols are the same as Figure \ref{fig:fitting_Ntop_fittingparameters_N2nd_model}. Red lines mark analytic formulae given by fitting relations of fitting parameters, derived from Equation (\ref{equ:fitting_AELandBEL_model}).}
  \label{fig:fitting_Ntop_fittingparameters_E1a2_model}
\end{figure}

Figure \ref{fig:fitting_Ntop_fittingparameters_E1a2_model} presents the parameters of Equation (\ref{equ:fitting_E1a2oNpro}) on $N_{\rm tar} / N_{\rm pro}$, which are fitted via
\begin{eqnarray}
  \begin{array}{l}
    A_{\rm EL} = 1.77 \times 10^{-9} \left ( \displaystyle \frac{N_{\rm tar}}{N_{\rm pro}} \right ) ^{-1.39} \mathrm{,} \\
    B_{\rm EL} = 12.4                \left ( \displaystyle \frac{N_{\rm tar}}{N_{\rm pro}} \right ) ^{-0.0454} \mathrm{,} \\
    A_{\rm EM} = 0.324 \left ( \displaystyle \frac{N_{\rm tar}}{N_{\rm pro}} \right ) ^{-0.838}  + 5.94 \times 10^{-6} \left ( \displaystyle \frac{N_{\rm tar}}{N_{\rm pro}} \right ) ^{5.17} \mathrm{,} \\
    B_{\rm EM} = -9.50 \left ( \displaystyle \frac{N_{\rm tar}}{N_{\rm pro}} \right ) ^{-0.0514} - 2.36 \times 10^{-2} \left ( \displaystyle \frac{N_{\rm tar}}{N_{\rm pro}} \right ) ^{1.84} + 9.46 \mathrm{,} \\
    A_{\rm EH} = \left \{ \left [ 2.82 \times 10^{-3} \left ( \displaystyle \frac{N_{\rm tar}}{N_{\rm pro}} \right ) ^{0.226} \right ] ^{-1} + \left [ 0.326 \left ( \displaystyle \frac{N_{\rm tar}}{N_{\rm pro}} \right ) ^{-1.04} \right ] ^{-1} \right \} ^{-1} \mathrm{,} \\
    \mathrm{and} \\
    B_{\rm EH} =                  1.12                \left ( \displaystyle \frac{N_{\rm tar}}{N_{\rm pro}} \right ) ^{0.0834} \mathrm{.}
  \end{array}
  \label{equ:fitting_AELandBEL_model}
\end{eqnarray}
Similarly to the mass of the second remnant (Equation (\ref{equ:fitting_N2ndoNpro_model})), we modify Equation (\ref{equ:fitting_E1a2oNpro}) with a cap $\langle M_{\rm eje} \rangle = M_{\rm tot} - (\langle M_{\rm lar} \rangle + \langle M_{\rm 2nd} \rangle) \le M_{\rm tot} - 2 \langle M_{\rm 2nd} \rangle$ by
\begin{equation}
  \frac{\langle M_{\rm eje} \rangle}{M_{\rm pro}} = \mathrm{min} \left ( \left \{ \left [ A_{\rm EL} \left ( \frac{v_{\rm col}}{v_{\rm bre}} \right ) ^{B_{\rm EL}} \right ] ^{-1} + \left [ A_{\rm EM} \left ( \frac{v_{\rm col}}{v_{\rm bre}} \right ) ^{B_{\rm EM}} + A_{\rm EH} \left ( \frac{v_{\rm col}}{v_{\rm bre}} \right ) ^{B_{\rm EH}} \right ] ^{-1} \right \} ^{-1} , \frac{M_{\rm tot} - 2 \langle M_{\rm 2nd} \rangle}{M_{\rm pro}} \right )
  \mathrm{.}
  \label{equ:fitting_E1a2oNpro_model}
\end{equation}
The fitting results of Equations (\ref{equ:fitting_AELandBEL_model}) and (\ref{equ:fitting_E1a2oNpro_model}) are shown in Figure \ref{fig:fitting_vcob_E1a2oNpro_Ntar_Ntop_ylog_fit_model}.
Figure \ref{fig:fitting_vcob_E1a2oNpro_Ntar_Ntop_ylog_fit_model} suggests that the analytic formulae of the total other ejecta mass can roughly reproduce the results of our simulations.

\begin{figure}
  \plotone{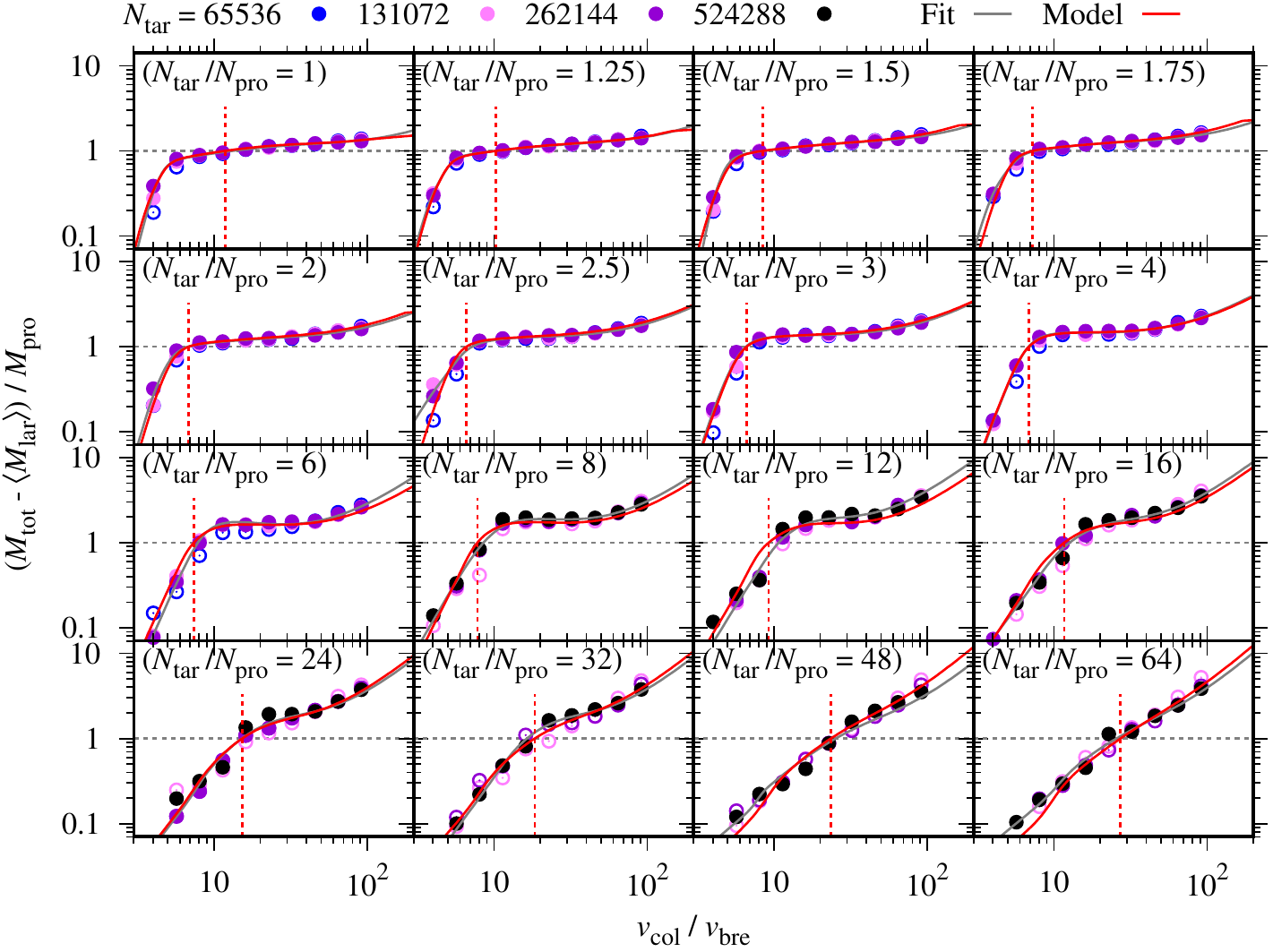}
  \caption{Same as Figure \ref{fig:fitting_vcob_E1a2oNpro_Ntar_Ntop_ylog_fit_model}, but for the fragments except for the largest remnant. Gray solid lines mark fitting relations derived from Equations (\ref{equ:fitting_N2ndoNpro}) and (\ref{equ:fitting_E1a2oNpro}), with assuming $B_{\rm SH} = -0.529$ and $B_{\rm EH} = 1.12 (N_{\rm tar} / N_{\rm pro})^{0.0834}$. Red solid lines indicate analytic formulae derived from Equations (\ref{equ:fitting_ASLandBSL_model}), (\ref{equ:fitting_N2ndoNpro_model}), (\ref{equ:fitting_AELandBEL_model}) and (\ref{equ:fitting_E1a2oNpro_model}). Horizontal gray dotted lines denote $\langle M_{\rm lar} \rangle = M_{\rm tar}$. Vertical red dotted lines represent $v_{\rm col} = v_{\rm fra}$ obtained from our fragment model. The red solid line and gray and red dotted lines always intersect at a point in each panel.}
  \label{fig:fitting_vcob_E1stoNpro_Ntar_Ntop_ylog_fit_model}
\end{figure}

The total-fragment mass except for the largest remnant can be derived from Equations (\ref{equ:fitting_ASLandBSL_model}), (\ref{equ:fitting_N2ndoNpro_model}), (\ref{equ:fitting_AELandBEL_model}) and (\ref{equ:fitting_E1a2oNpro_model}), $(M_{\rm tot} - \langle M_{\rm lar} \rangle) / M_{\rm pro} = \langle M_{\rm 2nd} \rangle / M_{\rm pro} + [M_{\rm tot} - (\langle M_{\rm lar} \rangle + \langle M_{\rm 2nd} \rangle)] / M_{\rm pro} = \langle M_{\rm 2nd} \rangle / M_{\rm pro} + \langle M_{\rm eje} \rangle / M_{\rm pro}$.
Figure \ref{fig:fitting_vcob_E1stoNpro_Ntar_Ntop_ylog_fit_model} shows the $b_{\mathrm{off}}$-weighted average total mass of the fragments except for the largest remnant, normalized by the projectile mass.
Figure \ref{fig:fitting_vcob_E1stoNpro_Ntar_Ntop_ylog_fit_model} suggests that our fragment model of the total-fragment mass except for the largest remnant can also nicely reproduce the simulation results.
In particular, the fitting formulae give reasonably good estimates of the critical collisional fragmentation velocity, which is the location where the solid line crosses the dotted line in each panel of Figure \ref{fig:fitting_vcob_E1stoNpro_Ntar_Ntop_ylog_fit_model}.

Figure \ref{fig:fitting_vcob_E1stoNpro_Ntar_Ntop_ylog_fit_model} shows that in collisions with high-velocity, $v_{\rm col} \gtrsim 2 \times 10^2 ~ \mathrm{m ~ s^{-1}}$, and high-mass ratio, $N_{\rm tar} / N_{\rm pro} \ge 12$, the power-law index of the remnant mass is about unity, $(M_{\rm tot} - \langle M_{\rm lar} \rangle) / M_{\rm pro} \propto v_{\rm col}$.
This dependence for higher-velocity collisions is steeper than that obtained for lower-velocity collisions as shown in Figure \ref{fig:fitting_vcob_E1stoNpro_Ntar_Ntop_ylog_fit_model} (see also Paper I for $v_{\rm col} \le 174 ~ \mathrm{m ~ s^{-1}}$).
On the other hand, Wada et al. (2013) reported that the proportional trend continues even for lower $v_{\rm col} < 2 \times 10^2 ~ \mathrm{m ~ s^{-1}}$ from their simulations with $N_{\rm tar} / N_{\rm pro} = 16$.
This is because the numerical resolution, $N_{\rm pro} =$ 2000 and 8000, adopted in Wada et al. (2013) is lower than our setup of $N_{\rm pro} =$ 16384 and 32768.
In simulations with a small projectile, the mass transfer from the target to the projectile is not well captured because of the insufficient numerical resolution (Figure \ref{fig:fitting_vcob_N2ndoNpro_Ntar_Ntop_ylin_fit_model}).
As a result, the $v_{\rm col}$ distribution of the remnants in Wada et al. (2013) does not have the component with a shallow slope in the middle velocity range, unlike shown in Figure \ref{fig:fitting_vcob_E1stoNpro_Ntar_Ntop_ylog_fit_model}.

\subsection{Mass Distributions of Small Fragments} \label{sec:res_small}

\begin{figure}
  \plotone{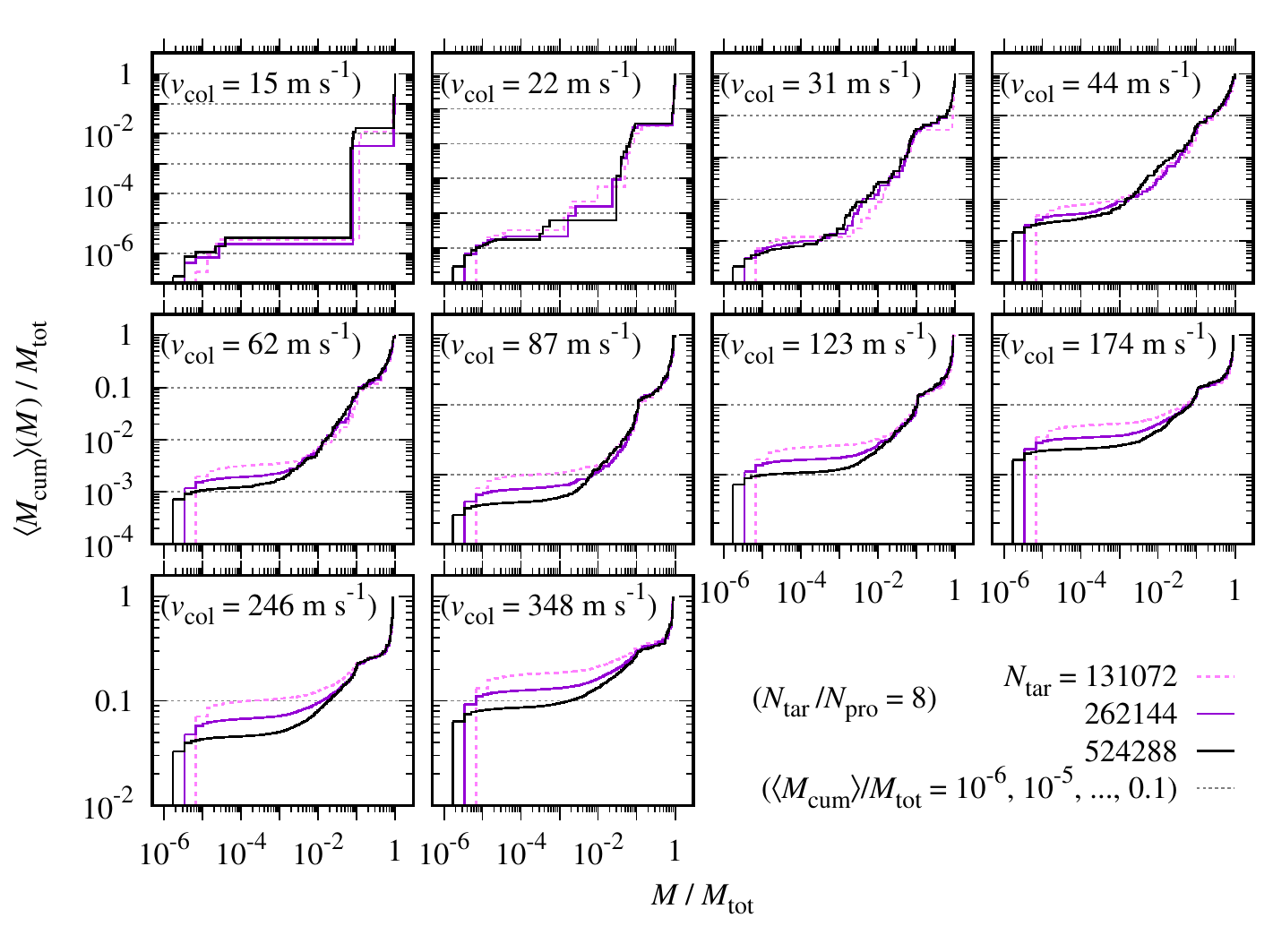}
  \caption{Cumulative mass distribution of fragments not larger than a given mass $M$, normalized by the total mass, $\langle M_{\rm cum} \rangle (M) / M_{\rm tot}$, for $N_{\rm tar} / N_{\rm pro} = 8$ (lines except for gray). Colors represent $N_{\rm tar} =$ 65,536 (blue), 131,072 (magenta), 262,144 (purple), and 524,288 (black), respectively. Horizontal gray dotted lines mark $\langle M_{\rm cum} \rangle / M_{\rm tot} = 10^{-6}, ~ 10^{-5}, ~ \ldots$, and 0.1. Different panels correspond to results of different collision velocities.}
  \label{fig:NoNt_Ncum_Ntar_vcol_Ntop8}
\end{figure}

The mass distribution of fragments can be written in the following cumulative form.
The distribution of the masses of fragments not larger than a given mass $M$ is defined as
\begin{equation}
  M_{\rm cum}(M) \equiv \int _{m_{\rm mon}}^M M n(M) dM
  \mathrm{,}
  \label{equ:Ncumulative}
\end{equation}
where $n(M) dM$ is the number of fragments with a mass between $M$ and $M + dM$.
Figure \ref{fig:NoNt_Ncum_Ntar_vcol_Ntop8} shows the cumulative mass distribution of the fragments, normalized by the total mass, $\langle M_{\rm cum} \rangle (M) / M_{\rm tot}$, for different collision velocities between colliding bodies with $N_{\rm tar} / N_{\rm pro} = 8$.
For fixed $N_{\rm tar} / N_{\rm pro}$, it is suggested that the cumulative mass distribution of relatively large fragments is independent of the total mass, while the distribution of relatively small fragments depends on it and the mass fraction of those small fragments tends to decrease with increasing the total mass.

Since tiny dust aggregates have a short growth timescale because of the Brownian motion in protoplanetary disks, the distribution of those tiny fragments ejected by a collision will change rapidly.
We neglect the detailed profile of the cumulative mass distribution of tiny fragments with $M \lesssim 10 m_{\rm mon}$ and assume that fragments except for the two large remnants are classified into fragments with the power-law mass distribution and monomer fragments as shown in Figure \ref{fig:diagram}.
We assume the fitting formula of the normalized cumulative mass distribution of fragments, in the mass range of fragments except for the two large remnants, as
\begin{equation}
  \frac{\langle M_{\rm cum} \rangle (M)}{M_{\rm tot}} = A_{\rm cum} \left ( \frac{M}{M_{\rm tot}} \right ) ^{B_{\rm cum}} + C_{\rm cum}
  \mathrm{,}
  \label{equ:fitting_NcumoNtot}
\end{equation}
where $A_{\rm cum}$, $B_{\rm cum}$, and $C_{\rm cum}$ are positive and are independent of the fragment mass $M$.
Because of $m_{\rm mon} \ll M_{\rm tot}$, $C_{\rm cum} M_{\rm tot} \equiv M_{\rm mon} \approx \langle M_{\rm cum} \rangle (m_{\rm mon})$ is the total mass of fragments in the individual-monomer state, i.e., $C_{\rm cum}$ is the mass fraction of monomer fragments, normalized by the total mass.
Neglecting the detailed profile for $M \lesssim 10 m_{\rm mon}$ means that $\langle M_{\rm cum} \rangle (10 m_{\rm mon}) \approx M_{\rm mon} = C_{\rm cum} M_{\rm tot}$; the fitting formula (\ref{equ:fitting_NcumoNtot}) overestimates the total mass of monomer fragments.

\begin{figure}
  \plotone{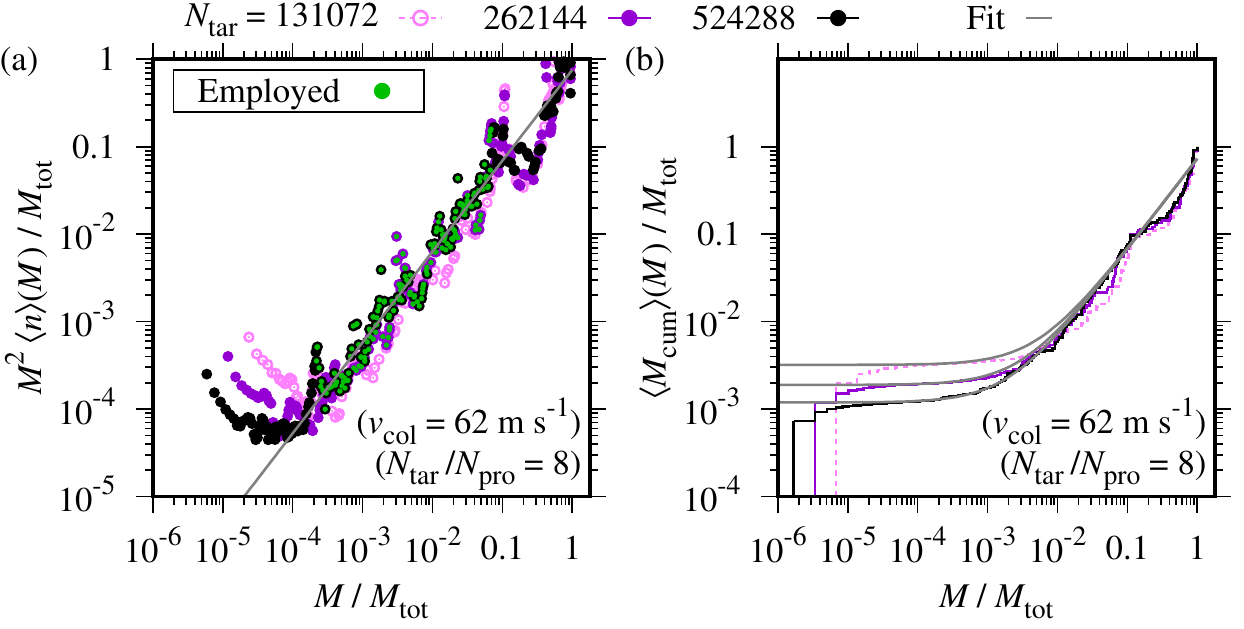}
  \caption{(a) Derivative of cumulative mass distribution of fragments, multiplied by $M / M_{\rm tot}$, $(M / M_{\rm tot}) \times d \langle M_{\rm cum} \rangle / dM = M^2 \langle n \rangle (M) / M_{\rm tot}$, for the case with $N_{\rm tar} / N_{\rm pro} = 8$ and $v_{\rm col} = 62 ~ \mathrm{m ~ s^{-1}}$ (circles). Green circles indicate data points employed in fits for $M \ge 128 m_{\rm mon}$ and $\langle M_{\rm cum} \rangle (M) \le \langle M_{\rm eje} \rangle = M_{\rm tot} - (\langle M_{\rm lar} \rangle + \langle M_{\rm 2nd} \rangle)$. The gray solid line marks the derivative of the fitting relation, $M^2 \langle n \rangle (M) / M_{\rm tot} = A_{\rm cum} B_{\rm cum} (M / M_{\rm tot})^{B_{\rm cum}}$. (b) Normalized cumulative mass distribution of fragments, $\langle M_{\rm cum} \rangle (M) / M_{\rm tot}$. Gray solid lines indicate fitting relations derived from Equation (\ref{equ:fitting_NcumoNtot}), $\langle M_{\rm cum} \rangle (M) / M_{\rm tot} = A_{\rm cum} (M / M_{\rm tot})^{B_{\rm cum}} + C_{\rm cum}$.}
  \label{fig:fitting_NoNt_cumlative_Ntar_v5_Ntop8}
\end{figure}

First, we determine the power-law component (first term) of Equation (\ref{equ:fitting_NcumoNtot}), because the intercept $C_{\rm cum}$ depends on the total mass $M_{\rm tot}$.
To do so, we take the derivative of Equation (\ref{equ:fitting_NcumoNtot}):
\begin{equation}
  \frac{d \langle M_{\rm cum} \rangle}{dM} = M \langle n \rangle (M) = A_{\rm cum} B_{\rm cum} \left ( \frac{M}{M_{\rm tot}} \right ) ^{B_{\rm cum} - 1}
  \label{equ:fitting_dNcumodNtot}
\end{equation}
(Figure \ref{fig:fitting_NoNt_cumlative_Ntar_v5_Ntop8}(a)).
From Figure \ref{fig:NoNt_Ncum_Ntar_vcol_Ntop8}, we adopt the range of the power-law component between $M \ge 128 m_{\rm mon}$ and $\langle M_{\rm cum} \rangle (M) \le \langle M_{\rm eje} \rangle = M_{\rm tot} - (\langle M_{\rm lar} \rangle + \langle M_{\rm 2nd} \rangle)$ for each $N_{\rm tar}$.
The upper bound of the power-law component, $M_{\rm pow}$, is determined by $\langle M_{\rm cum} \rangle (M_{\rm pow}) = \langle M_{\rm eje} \rangle$.
Then we obtain
\begin{equation}
  \frac{M_{\rm pow}}{M_{\rm tot}} = \left \{ \frac{1}{A_{\rm cum}} \left [ \left ( \frac{N_{\rm tar}}{N_{\rm pro}} + 1 \right ) ^{-1} \frac{\langle M_{\rm eje} \rangle}{M_{\rm pro}} - C_{\rm cum} \right ] \right \} ^{1 / B_{\rm cum}}
  \mathrm{.}
  \label{equ:Nmpd}
\end{equation}
For $v_{\rm col} \ge 174 ~ \mathrm{m ~ s^{-1}}$, the additional restriction on the upper bound, $M \le 3 \times 10^{-3} M_{\rm tot}$, is also applied (see Section \ref{sec:dis_com_lab} for the reason).

Second, we determine the dependence of the intercept $C_{\rm cum} = M_{\rm mon} / M_{\rm tot}$ on the total mass in Equation (\ref{equ:fitting_NcumoNtot}) with employing $A_{\rm cum}$ and $B_{\rm cum}$ obtained above (Figure \ref{fig:fitting_NoNt_cumlative_Ntar_v5_Ntop8}(b)), where we assume the range of the mass for fragments used in fits to be $M \ge 10 m_{\rm mon}$, as discussed previously.
The slopes of the gray lines in both panels of Figure \ref{fig:fitting_NoNt_cumlative_Ntar_v5_Ntop8} are $B_{\rm cum}$, because $M^2 \langle n \rangle (M) / M_{\rm tot} = A_{\rm cum} B_{\rm cum} (M / M_{\rm tot})^{B_{\rm cum}} \propto M^{B_{\rm cum}}$ (Figure \ref{fig:fitting_NoNt_cumlative_Ntar_v5_Ntop8}(a)) and $\langle M_{\rm cum} \rangle (M) / M_{\rm tot} \propto M^{B_{\rm cum}}$ for $\langle M_{\rm cum} \rangle (M) / M_{\rm tot} \gg C_{\rm cum}$ (Figure \ref{fig:fitting_NoNt_cumlative_Ntar_v5_Ntop8}(b)), whereas the slope of the cumulative distribution $\langle M_{\rm cum} \rangle (M)$ in the small mass range $M \lesssim 10^{-3} M_{\rm tot}$ is shallower than $B_{\rm cum}$ because of the effect of the single monomers $C_{\rm cum}$.

\begin{figure}
  \plotone{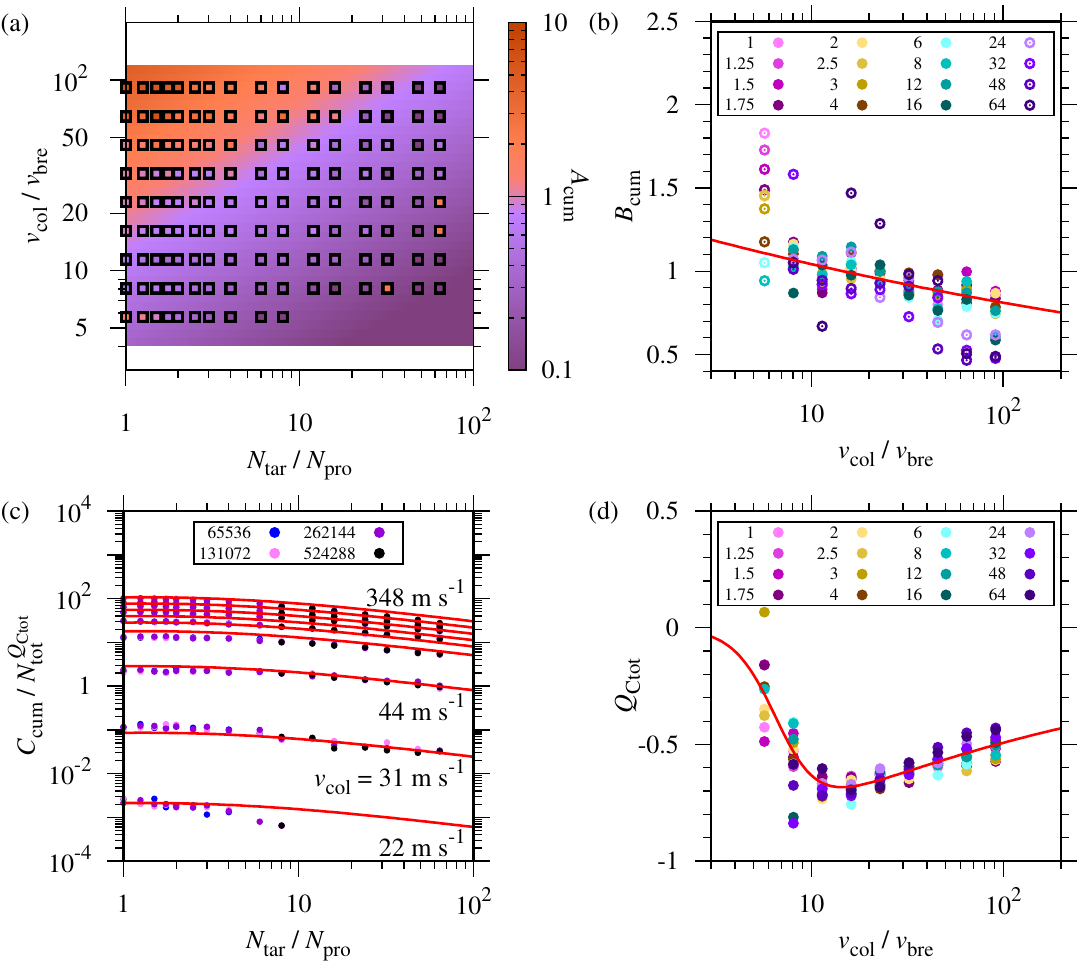}
  \caption{(a) Coefficient of the fitting formula (\ref{equ:fitting_NcumoNtot}), $A_{\rm cum}$, on the $N_{\rm tar} / N_{\rm pro}$-$v_{\rm col} / v_{\rm bre}$ plane (colors and squares). The squares indicate the sets of our simulations. Colors inside and outside squares represent our results of simulations and the fitting relation derived from Equation (\ref{equ:fitting_Acum}), respectively. (b) Power-law index of the fitting formula (\ref{equ:fitting_NcumoNtot}), $B_{\rm cum}$, against the normalized collision velocity, $v_{\rm col} / v_{\rm bre}$ (circles). Colors represent the mass ratios of colliding bodies, $N_{\rm tar} / N_{\rm pro}$. Filled and open circles are the same as Figure \ref{fig:fitting_vcob_N2ndoNpro_Ntar_Ntop_ylin_fit_model}. The red solid line marks the fitting relation derived from Equation (\ref{equ:fitting_Bcum}). (c) Intercept of the fitting formula (\ref{equ:fitting_NcumoNtot}), normalized by $N_{\rm tot}^{Q_{\rm Ctot}}$, $C_{\rm cum} / N_{\rm tot}^{Q_{\rm Ctot}}$, for the initial mass ratio, $N_{\rm tar} / N_{\rm pro}$ (circles and lines except for the red lines). Colors represent the number of dust monomers of the target, $N_{\rm tar}$. Red solid lines indicate fitting relations derived from Equations (\ref{equ:fitting_Ccum}) and (\ref{equ:fitting_QCtot}). Different red lines correspond to different collision velocities of $v_{\rm col} =$ 22 $\mathrm{m ~ s^{-1}}$, 31 $\mathrm{m ~ s^{-1}}$, 44 $\mathrm{m ~ s^{-1}}$, 62 $\mathrm{m ~ s^{-1}}$, 87 $\mathrm{m ~ s^{-1}}$, 123 $\mathrm{m ~ s^{-1}}$, 174 $\mathrm{m ~ s^{-1}}$, 246 $\mathrm{m ~ s^{-1}}$, and 348 $\mathrm{m ~ s^{-1}}$ in ascending order of the vertical axis $C_{\rm cum} / N_{\rm tot}^{Q_{\rm Ctot}}$. (d) Dependence of the fitting parameter $C_{\rm cum}$ on the total monomer number, $Q_{\rm Ctot}$, against the normalized collision velocity, $v_{\rm col} / v_{\rm bre}$ (circles). Colors represent the mass ratios of colliding bodies, $N_{\rm tar} / N_{\rm pro}$. The red line denotes the fitting relation derived from Equation (\ref{equ:fitting_QCtot}).}
  \label{fig:fitting_Ntop_vcob_fittingABC_Ncum}
\end{figure}

Finally, we investigate the dependence of the obtained $A_{\rm cum}$, $B_{\rm cum}$, and $C_{\rm cum}$ on $N_{\rm tot}$, $N_{\rm tar} / N_{\rm pro}$ and $v_{\rm col}$.
Collisions with low-mass ratio, $N_{\rm tar} / N_{\rm pro} \le 16$, are employed for the fits of $A_{\rm cum}$ and $B_{\rm cum}$ because there is an insufficient power-law range in higher mass-ratio collisions.
In fits of $C_{\rm cum}$, results of simulations with $N_{\rm tar} \ge$ 65536 and 131072 are also used for $N_{\rm tar} / N_{\rm pro} =$ 1 to 6 and 8 to 64, respectively.
Figure \ref{fig:fitting_Ntop_vcob_fittingABC_Ncum} shows the fitting parameters given by
\begin{equation}
  A_{\rm cum} = 0.108 \left ( \frac{N_{\rm tar}}{N_{\rm pro}} \right ) ^{-0.416} \left ( \frac{v_{\rm col}}{v_{\rm bre}} \right ) ^{0.790}
  \mathrm{,}
  \label{equ:fitting_Acum}
\end{equation}
\begin{equation}
  B_{\rm cum} = 1.34 \left ( \frac{v_{\rm col}}{v_{\rm bre}} \right ) ^{-0.109}
  \mathrm{,}
  \label{equ:fitting_Bcum}
\end{equation}
and
\begin{eqnarray}
  \begin{array}{rl}
    C_{\rm cum} = N_{\rm tot}^{Q_{\rm Ctot}} & \left \{ \left [ \left ( \frac{N_{\rm tar}}{N_{\rm pro}} \right ) ^{0.296} \right ] ^{-1} + \left [ 2.09 \left ( \frac{N_{\rm tar}}{N_{\rm pro}} \right ) ^{-0.508} \right ] ^{-1} \right \} ^{-1} \\
    { } & \times \left \{ \left [ 2.56 \times 10^{-11} \left ( \frac{v_{\rm col}}{v_{\rm bre}} \right ) ^{10.7} \right ] ^{-1} + \left [ 2.14 \left ( \frac{v_{\rm col}}{v_{\rm bre}} \right ) ^{0.953} \right ] ^{-1} \right \} ^{-1}
  \end{array}
  \mathrm{,}
  \label{equ:fitting_Ccum}
\end{eqnarray}
where
\begin{equation}
  Q_{\rm Ctot} = - \left \{ \left [ 6.34 \times 10^{-4} \left ( \frac{v_{\rm col}}{v_{\rm bre}} \right ) ^{3.74} \right ] ^{-1} + \left [ 1.21 \left ( \frac{v_{\rm col}}{v_{\rm bre}} \right ) ^{-0.195} \right ] ^{-1} \right \} ^{-1}
  \mathrm{.}
  \label{equ:fitting_QCtot}
\end{equation}
The coefficient $A_{\rm cum}$ corresponds to the mass fraction of the fragments in the power-law range (green region in Figure \ref{fig:diagram}).
Since for collisions with high-mass ratio the projectile is much smaller than the target, the fraction of the second-largest remnants is also small, and therefore $A_{\rm cum}$ is smaller for higher $N_{\rm tar} / N_{\rm pro}$ (Figure \ref{fig:fitting_Ntop_vcob_fittingABC_Ncum}(a)).
The coefficient $A_{\rm cum}$ increases with $v_{\rm col}$ because the fraction of the small fragments is larger for high-$v_{\rm col}$ collisions.
Figure \ref{fig:fitting_Ntop_vcob_fittingABC_Ncum}(b) shows that the power-law index of the cumulative mass distribution, $B_{\rm cum}$, is almost independent of the mass ratio, which indicates that $B_{\rm cum}$ is not affected by the mass transfer from a target to a projectile.
In contrast, $B_{\rm cum}$ depends on the collision velocity.
For the low collision velocity, the power-law index is steep because the fragments are divided into a few large remnants and a large amount of tiny fragments.
For the high collision velocity, on the other hand, the cumulative mass distribution gradually increases because the fragments with various sizes are ejected.
Thus, $B_{\rm cum}$ decreases with increasing the collision velocity $v_{\rm col}$.

The mass fraction of monomer fragments, $C_{\rm cum}$, depends weakly on the mass ratio and strongly on the collision velocity.
The dependence of $M_{\rm mon} = C_{\rm cum} M_{\rm tot}$ on $M_{\rm tot}$ is given by $M_{\rm mon} \propto M_{\rm tot}^{Q_{\rm Ctot} + 1}$, and Figure \ref{fig:fitting_Ntop_vcob_fittingABC_Ncum}(d) indicates $0 < Q_{\rm Ctot} + 1 \le 1$.
Thus, for a fixed mass ratio, while the mass of monomer fragments, $M_{\rm mon}$, gradually increases with the projectile mass, the mass fraction of monomer fragments, $C_{\rm cum}$, gradually decreases with increasing the projectile mass.
In order to create many individual dust monomers by collisions, not just once but multiple hierarchical disruptive collisions (i.e., collisions between fragments, collisions between fragments of fragments) are required in realistic situations.
Some previous works consider the case when all fragments are always divided into individual dust monomers (e.g., regime A in Birnstiel et al. 2011; Homma et al. 2019).
However, our results imply that such complete fragmentation hardly occurs.
While Paszun \& Dominik (2009) showed that many tiny fragments are ejected for high-velocity collisions, they adopted initial aggregates composed of $10^3$ particles.
Osinsky \& Brilliantov (2022) also showed that the fragment mass distribution is steep and many tiny fragments are ejected for high-velocity collisions.
However, they adopted initial aggregates composed of $\lesssim 10^4$ particles and 
did not take into account the tangential motions (rolling, sliding, and twisting), while they performed $\sim 10^3$-$10^4$ runs for each set of parameters.

\begin{figure}
  \plotone{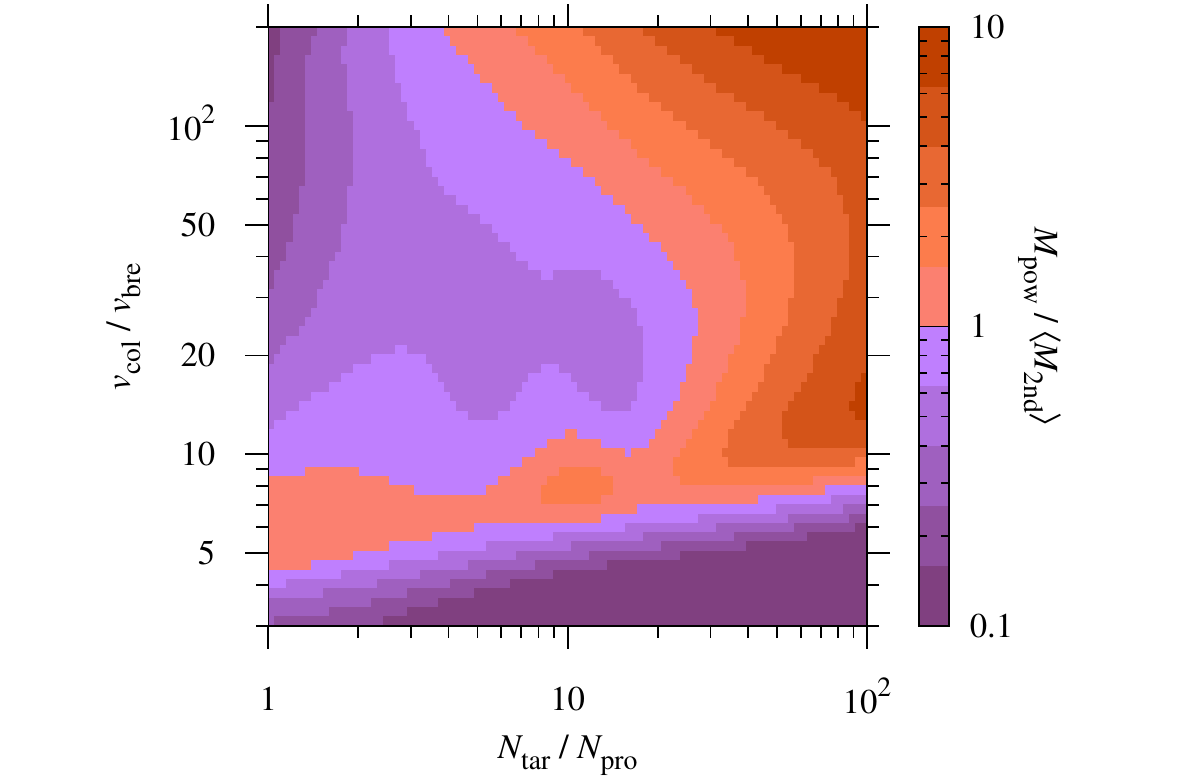}
  \caption{Mass ratio $M_{\rm pow} / \langle M_{\rm 2nd} \rangle$ on the $N_{\rm tar} / N_{\rm pro}$-$v_{\rm col} / v_{\rm bre}$ plane. The upper limit of the fragment mass with the cumulative mass distribution of fragments not larger than the total other ejecta mass, $M_{\rm pow}$, is derived from Equation (\ref{equ:Nmpd}). The $b_{\mathrm{off}}$-weighted average mass of the second-largest remnant, $\langle M_{\rm 2nd} \rangle$, is derived from Equation (\ref{equ:fitting_N2ndoNpro_model}).}
  \label{fig:fitting_Ntop_vcob_NmpdoN2nd}
\end{figure}

Using Equations (\ref{equ:fitting_NcumoNtot})-(\ref{equ:fitting_QCtot}), we can calculate the mass distribution of fragments in the cumulative form from the total mass, the mass ratio, and the collision velocity between colliding bodies.
However, there is a caveat for collisions with high-mass ratio and high collision velocity.
In such collisions, $\langle M_{\rm 2nd} \rangle$ derived from the numerical fitting (Equation (\ref{equ:fitting_N2ndoNpro_model})) could be smaller than $M_{\rm pow}$ estimated by Equation (\ref{equ:Nmpd}) (Figure \ref{fig:fitting_Ntop_vcob_NmpdoN2nd}).
In other words, because only the largest remnant is much larger than other fragments, the second remnant has a size within the power-law mass distribution.
However, $\langle M_{\rm cum} \rangle (M = \langle M_{\rm 2nd} \rangle)$ in Equation (\ref{equ:fitting_NcumoNtot}) should not be smaller than $\langle M_{\rm eje} \rangle$ for the consistency of the model construction, because the other ejecta do not contain the two large remnants.
In order to construct a self-consistent fragment model, we artificially modify the fitting formula of the normalized cumulative mass distribution of fragments by
\begin{equation}
  \frac{\langle M_{\rm cum} \rangle (M)}{M_{\rm tot}} = A_{\rm cum} \mathrm{max} \left [ 1, \left ( \frac{M_{\rm pow}}{\langle M_{\rm 2nd} \rangle} \right ) ^{B_{\rm cum}} \right ] \left ( \frac{M}{M_{\rm tot}} \right ) ^{B_{\rm cum}} + C_{\rm cum}
  \mathrm{,}
  \label{equ:fitting_NcumoNtot_model}
\end{equation}
where $\mathrm{max} [1, (M_{\rm pow} / \langle M_{\rm 2nd} \rangle)^{B_{\rm cum}}]$ ensures that the mass of the third-largest remnant is certainly smaller than the mass of the second remnant, and $M_{\rm pow} / \langle M_{\rm 2nd} \rangle = (1 + N_{\rm tar} / N_{\rm pro}) (M_{\rm pow} / M_{\rm tot}) (\langle M_{\rm 2nd} \rangle / M_{\rm pro})^{-1}$.
Equation (\ref{equ:fitting_NcumoNtot_model}) with Equations (\ref{equ:Nmpd})-(\ref{equ:fitting_QCtot}) gives the cumulative mass distribution of fragments from the total mass, the mass ratio, and the collision velocity between colliding bodies.

When $\langle M_{\rm eje} \rangle \le M_{\rm mon} = C_{\rm cum} M_{\rm tot}$, all fragments except for the two large remnants are divided into individual dust monomers, while such a collision will not occur in realistic collisional fragmentation except for collisions with low collision velocities (Section \ref{sec:dis_fra_eje}).

Equations (\ref{equ:fitting_N2ndoNpro_model}), (\ref{equ:fitting_E1a2oNpro_model}), and (\ref{equ:fitting_NcumoNtot_model}) with Equations (\ref{equ:fitting_ASLandBSL_model}), (\ref{equ:fitting_AELandBEL_model}), and (\ref{equ:Nmpd})-(\ref{equ:fitting_QCtot}) gives the mass distribution of all fragments from the total mass, the mass ratio, and the collision velocity between colliding bodies (Table \ref{tab:procedure} and Appendix \ref{sec:app_res}).
More specifically, $\langle M_{\rm lar} \rangle$, $\langle M_{\rm 2nd} \rangle$ (Equation (\ref{equ:fitting_N2ndoNpro_model})), $\langle M_{\rm eje} \rangle$ (Equation (\ref{equ:fitting_E1a2oNpro_model})), and $\langle M_{\rm cum} \rangle$ (Equation (\ref{equ:fitting_NcumoNtot_model})) are derived from the input parameters of $N_{\rm tot}$, $N_{\rm tar} / N_{\rm pro}$, and $v_{\rm col} / v_{\rm bre}$.
Here we note that $\langle M_{\rm lar} \rangle$, $\langle M_{\rm 2nd} \rangle$, and $\langle M_{\rm eje} \rangle$ are not independent each other (Equations (\ref{equ:fitting_N2ndoNpro_model}) and (\ref{equ:fitting_E1a2oNpro_model})), and therefore, the three output variables are derived from the three input parameters in our prescription.

\begin{table}
  \centering
  \caption{Procedure to Calculate the Mass Distribution of Fragments}
  \label{tab:procedure}
  \begin{tabular}{cc|l}
    \hline
    \hline
    Outputs & Inputs & Procedure to obtain left-mentioned variables \\
    \hline
    $\langle M_{\rm 2nd} \rangle$ & $N_{\rm tar} / N_{\rm pro}$, $v_{\rm col} / v_{\rm bre}$ & (1-1) Calculating $A_{\rm SL}$, $B_{\rm SL}$, $A_{\rm SM}$, $B_{\rm SM}$, $A_{\rm SH}$, and $B_{\rm SH}$ by Equation (\ref{equ:fitting_ASLandBSL_model}) with $N_{\rm tar} / N_{\rm pro}$. \\
    { } & { } & (1-2) Calculating $\langle M_{\rm 2nd} \rangle / M_{\rm pro}$ by Equation (\ref{equ:fitting_N2ndoNpro_model}) with $v_{\rm col} / v_{\rm bre}$ and coefficients and \\
    { } & { } & power-law indices derived in step 1-1. \\
    \hline
    $\langle M_{\rm eje} \rangle$ & $N_{\rm tar} / N_{\rm pro}$, $v_{\rm col} / v_{\rm bre}$, & (2-1) Calculating $A_{\rm EL}$, $B_{\rm EL}$, $A_{\rm EM}$, $B_{\rm EM}$, $A_{\rm EH}$, and $B_{\rm EH}$ by Equation (\ref{equ:fitting_AELandBEL_model}) with $N_{\rm tar} / N_{\rm pro}$. \\
    { } & $\langle M_{\rm 2nd} \rangle / M_{\rm pro}$ & (2-2) Calculating $\langle M_{\rm eje} \rangle / M_{\rm pro}$ by Equation (\ref{equ:fitting_E1a2oNpro_model}) with $v_{\rm col} / v_{\rm bre}$, $M_{\rm tot} / M_{\rm pro} = 1 + N_{\rm tar} / N_{\rm pro}$, \\
    { } & { } & $\langle M_{\rm 2nd} \rangle / M_{\rm pro}$ derived in step 1-2, and coefficients and power-law indices derived in step 2-1. \\
    \hline
    $\langle M_{\rm lar} \rangle$ & $N_{\rm tar} / N_{\rm pro}$, & (3-1) Calculating $(M_{\rm tot} - \langle M_{\rm lar} \rangle) / M_{\rm pro}$ by the sum of $\langle M_{\rm 2nd} \rangle / M_{\rm pro}$ and $\langle M_{\rm eje} \rangle / M_{\rm pro}$. \\
    { } & $\langle M_{\rm 2nd} \rangle / M_{\rm pro}$, $\langle M_{\rm eje} \rangle / M_{\rm pro}$ & (3-2) Calculating $\langle M_{\rm lar} \rangle / M_{\rm pro} = M_{\rm tot} / M_{\rm pro} - (M_{\rm tot} - \langle M_{\rm lar} \rangle) / M_{\rm pro}$ with \\
    { } & { } & $M_{\rm tot} / M_{\rm pro} = 1 + N_{\rm tar} / N_{\rm pro}$. \\
    \hline
    $\langle M_{\rm cum} \rangle$ & $N_{\rm tot}$, $N_{\rm tar} / N_{\rm pro}$, $v_{\rm col} / v_{\rm bre}$, & (4-1) Calculating $A_{\rm cum}$, $B_{\rm cum}$, $C_{\rm cum}$, and $M_{\rm pow} / M_{\rm tot}$ by Equations (\ref{equ:Nmpd})-(\ref{equ:fitting_QCtot}) with \\
    { } & $\langle M_{\rm 2nd} \rangle / M_{\rm pro}$, $\langle M_{\rm eje} \rangle / M_{\rm pro}$ & $v_{\rm col} / v_{\rm bre}$, $N_{\rm tar} / N_{\rm pro}$, $N_{\rm tot}$, and $\langle M_{\rm eje} \rangle / M_{\rm pro}$. \\
    { } & { } & (4-2-1) When $\langle M_{\rm eje} \rangle \le M_{\rm mon} = C_{\rm cum} M_{\rm tot}$, all fragments except for the two large \\
    { } & { } & remnants are divided into individual dust monomers. \\
    { } & { } & (4-2-2) When $\langle M_{\rm eje} \rangle > M_{\rm mon}$, calculating $\langle M_{\rm cum} \rangle / M_{\rm tot}$ by Equation (\ref{equ:fitting_NcumoNtot_model}) with \\
    { } & { } & $N_{\rm tar} / N_{\rm pro}$, $\langle M_{\rm 2nd} \rangle / M_{\rm pro}$, and $A_{\rm cum}$, $B_{\rm cum}$, $C_{\rm cum}$, and $M_{\rm pow} / M_{\rm tot}$ derived in step 4-1. \\
    \hline
  \end{tabular}
\end{table}

\section{Discussion} \label{sec:dis}

\subsection{Contribution of Monomer Fragments to "Other Ejecta"} \label{sec:dis_fra_eje}

In modeling, we classified the other ejecta except for the two large remnants into fragments with the power-law mass distribution and monomer fragments as shown in Figure \ref{fig:diagram}.
This is because the power-law component for low to middle collision velocities and low to middle mass ratios is much larger than the monomer component and is almost independent of the total mass, while the monomer component depends on the total mass.
However, inspecting our numerical results, we find that the power-law component is affected by $M_{\rm tot}$ for collisions with high $v_{\rm col}$ and large $N_{\rm tar} / N_{\rm pro}$ particularly when $N_{\rm tot} < 10^6$ (Appendix \ref{sec:app_res} and Figures \ref{fig:appendix_fitting_NoNt_Ncum_Ntar_vcol_Ntop1}-\ref{fig:appendix_fitting_NoNt_Ncum_Ntar_vcol_Ntop32}).
Figures in Appendix \ref{sec:app_res} suggest that the cause of this contribution of tiny fragments may be the insufficient numerical resolution, $N_{\rm tot} < 10^6$, since the contribution is more significant for the smaller $N_{\rm tot}$ and/or $N_{\rm pro}$.
Here we discuss the dependence of $M_{\rm mon}$ and $M_{\rm eje}$ on the adopted numerical resolution.
Because all dust monomers have the same mass, we discuss the monomer number constituting fragments instead of the fragment masses.

At the high collision velocity, the normalized total other ejecta mass, derived from Equations (\ref{equ:fitting_ASLandBSL_model}), (\ref{equ:fitting_N2ndoNpro_model}), (\ref{equ:fitting_AELandBEL_model}) and (\ref{equ:fitting_E1a2oNpro_model}), is given by $\langle M_{\rm eje} \rangle / M_{\rm pro} = \langle N_{\rm eje} \rangle / N_{\rm pro} \sim A_{\rm EH} (v_{\rm col} / v_{\rm bre})^{B_{\rm EH}}$.
From Figure \ref{fig:fitting_Ntop_fittingparameters_E1a2_model}, $A_{\rm EH} \sim 3 \times 10^{-3}$ and $B_{\rm EH} > 1$.
For the low mass ratio, the number of monomer fragments, derived from Equations (\ref{equ:fitting_Ccum}) and (\ref{equ:fitting_QCtot}), is roughly estimated to be $N_{\rm mon} = C_{\rm cum} N_{\rm tot} \sim N_{\rm tot}^{Q_{\rm Ctot} + 1} \times 2 (v_{\rm col} / v_{\rm bre})^1 \propto N_{\rm pro}^{Q_{\rm Ctot} + 1} v_{\rm col}^1$ because Figure \ref{fig:fitting_Ntop_vcob_fittingABC_Ncum}(c) shows that $C_{\rm cum}$ is roughly independent of $N_{\rm tar} / N_{\rm pro}$ for the low mass ratio.
Then we have $N_{\rm mon} / \langle N_{\rm eje} \rangle \propto N_{\rm pro}^{Q_{\rm Ctot}} v_{\rm col}^{1 - B_{\rm EH}}$.
Since $Q_{\rm Ctot} < 0$ and $B_{\rm EH} > 1$, $N_{\rm mon} \ll \langle N_{\rm eje} \rangle$ for the low mass ratio with $N_{\rm pro} \rightarrow \infty$ even for the high $v_{\rm col}$.
For the high mass ratio, on the other hand, $N_{\rm mon} \sim N_{\rm pro}^{0.5} \times 2 (v_{\rm col} / v_{\rm bre})^1$ because of $N_{\rm tot} \approx N_{\rm tar}$.
Then $N_{\rm mon} / \langle N_{\rm eje} \rangle \propto N_{\rm pro}^{-0.5} v_{\rm col}^{1 - B_{\rm EH}} \ll 1$ at $N_{\rm pro} \rightarrow \infty$ even for the high $v_{\rm col}$.
Thus, for realistic collisional fragmentation, the total mass of monomer fragments, $M_{\rm mon}$, is much smaller than the total other ejecta mass $\langle M_{\rm eje} \rangle$, and $\langle M_{\rm eje} \rangle$ is dominated by the fragments in the regime of the power-law distribution.

Although our fitting formula of $\langle M_{\rm eje} \rangle$ (Equation (\ref{equ:fitting_E1a2oNpro_model})) is independent of $N_{\rm pro}$, Figure \ref{fig:fitting_vcob_E1a2oNpro_Ntar_Ntop_ylog_fit_model} implies that the total other ejecta mass $\langle M_{\rm eje} \rangle$ obtained by our simulations is dependent on $N_{\rm pro}$ yet for cases with high mass ratios and/or high collision velocities.
On the other hand, our results also imply that $\langle M_{\rm eje} \rangle - \langle M_{\rm cum} \rangle (M = m_{\rm mon})$ seems to be roughly independent of $N_{\rm pro}$.
However, $\langle M_{\rm cum} \rangle (M)$ requires the information of $\langle M_{\rm eje} \rangle$ (Equations (\ref{equ:Nmpd}) and (\ref{equ:fitting_NcumoNtot_model})).
Therefore, we can not utilize $\langle M_{\rm eje} \rangle - \langle M_{\rm cum} \rangle (m_{\rm mon})$ for modeling.

\subsection{Cross Section of Dust Aggregates} \label{sec:dis_cro_sec}

The cross sections of dust aggregates for scattering and absorption of radiation are important to understand the observations of protoplanetary disks (Bohren \& Huffman 1983; Kataoka et al. 2014; Cuzzi et al. 2014; Okuzumi et al. 2016; Tazaki et al. 2019a, 2019b).
The electric charging of dust aggregates in protoplanetary disks is controlled by the collisional cross section between dust aggregates and charged particles (ions or electrons) (Okuzumi 2009).
To evaluate those effects, we estimate the geometric cross sections of the fragments resulting from collisions between dust aggregates.

Here we follow the simple method in Arakawa et al. (2022) and calculate the geometric cross sections of the fragments assumed to be spherical-shaped compact dust grains for simplicity.
We note that this assumption may underestimate the geometric cross sections of large fluffy fragments.
Under this assumption, the geometric cross section of a dust grain with $M = N m_{\rm mon}$ is given by $S(N) = \pi (0.1 N^{1/3} ~ \mathrm{\micron})^2$, because the fractal dimension is three and the mass is proportional to the monomer number.
Thus, the ratio of the cumulative geometric cross section of fragments not larger than a given mass $M$ to the geometric cross section of a dust grain with $M = M_{\rm tot}$ is given by
\begin{equation}
  S_{\rm cum} (M) = \displaystyle \frac{\displaystyle \int _{m_{\rm mon}}^M S(N) n(M) dM}{S(N_{\rm tot})} = \int _{m_{\rm mon}}^M \left ( \frac{M}{M_{\rm tot}} \right ) ^{2/3} n(M) dM
  \mathrm{,}
  \label{equ:Scumulative}
\end{equation}
where $N / N_{\rm tot} = M / M_{\rm tot}$ because of equal-mass monomers.
This ratio $S_{\rm cum}$ at $M = M_{\rm tot}$ is not smaller than unity, and $S_{\rm cum} = 1$ when $M_{\rm lar} = M_{\rm tot}$.

\begin{figure}
  \plotone{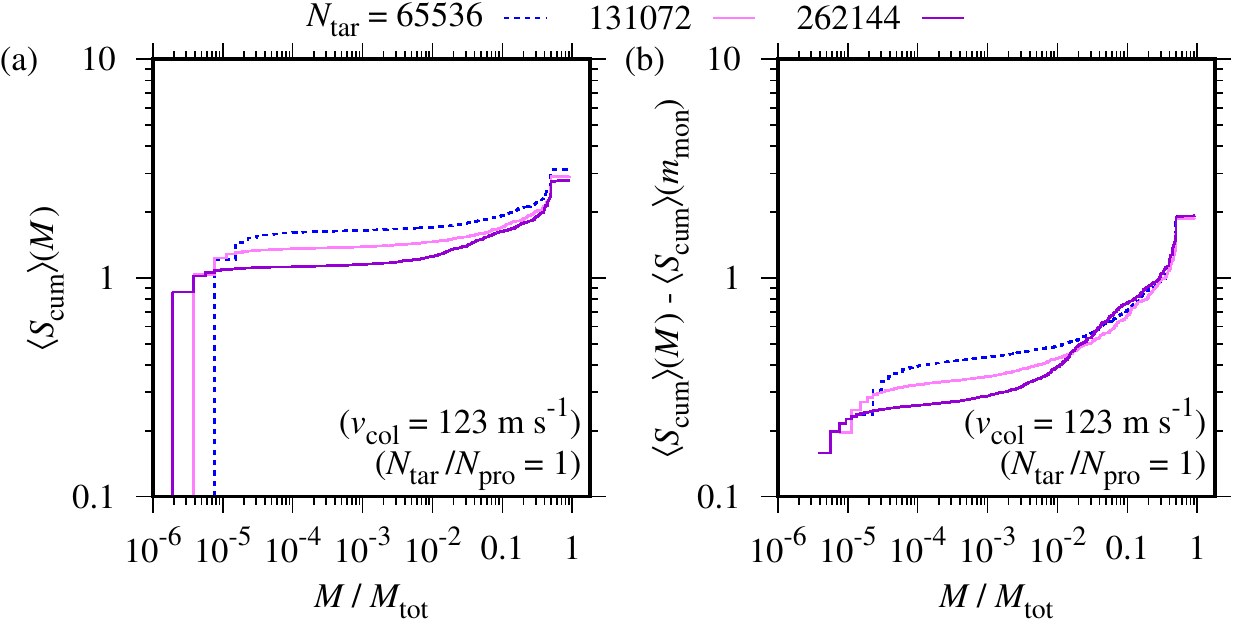}
  \caption{(a) Cumulative geometric cross section distribution of fragments not larger than a given mass $M$, normalized by the geometric cross section of a dust aggregate with $M = M_{\rm tot}$, $\langle S_{\rm cum} \rangle (M)$, for the case with $N_{\rm tar} / N_{\rm pro} = 1$ and $v_{\rm col} = 123 ~ \mathrm{m ~ s^{-1}}$, with assuming that all dust aggregates are spherical-shaped compact dust grains. Colors are the same as Figure \ref{fig:NoNt_Ncum_Ntar_vcol_Ntop8}. (b) Same as panel (a) but for the fragments except for monomer fragments, $\langle S_{\rm cum} \rangle (M) - \langle S_{\rm cum} \rangle (m_{\rm mon})$.}
  \label{fig:NoNt_cumlative_Ntar_v7_Ntop1}
\end{figure}

As shown in Section \ref{sec:dis_fra_eje}, the mass distribution of fragments except for small ones will be independent of $N_{\rm tot}$ even for high collision velocities if $N_{\rm pro} \rightarrow \infty$, since the total mass of monomer fragments hardly contributes to the total other ejecta mass.
However, as shown in Section \ref{sec:res_small} and Appendix \ref{sec:app_res}, results of our simulations with high collision velocities are affected by a large amount of tiny fragments owing to the insufficient numerical resolution, $N_{\rm tot} < 10^6$.
Our results show that both $\langle S_{\rm cum} \rangle (M = m_{\rm mon})$ and $\langle S_{\rm cum} \rangle (M = M_{\rm tot})$ decrease with increasing $N_{\rm tar}$ for fixed $N_{\rm tar} / N_{\rm pro}$ (Figure \ref{fig:NoNt_cumlative_Ntar_v7_Ntop1}(a)).
On the other hand, our results also show that $\langle S_{\rm cum} \rangle (M) - \langle S_{\rm cum} \rangle (m_{\rm mon})$ is roughly independent of $N_{\rm tar}$ for large fragments, whereas it decreases with increasing $N_{\rm tar}$ for tiny fragments (Figure \ref{fig:NoNt_cumlative_Ntar_v7_Ntop1}(b)).
Therefore, we focus on $\langle S_{\rm cum} \rangle (M) - \langle S_{\rm cum} \rangle (m_{\rm mon})$ and estimate the mass of fragments that dominate the cumulative geometric cross section of all fragments.

\begin{figure}
  \plotone{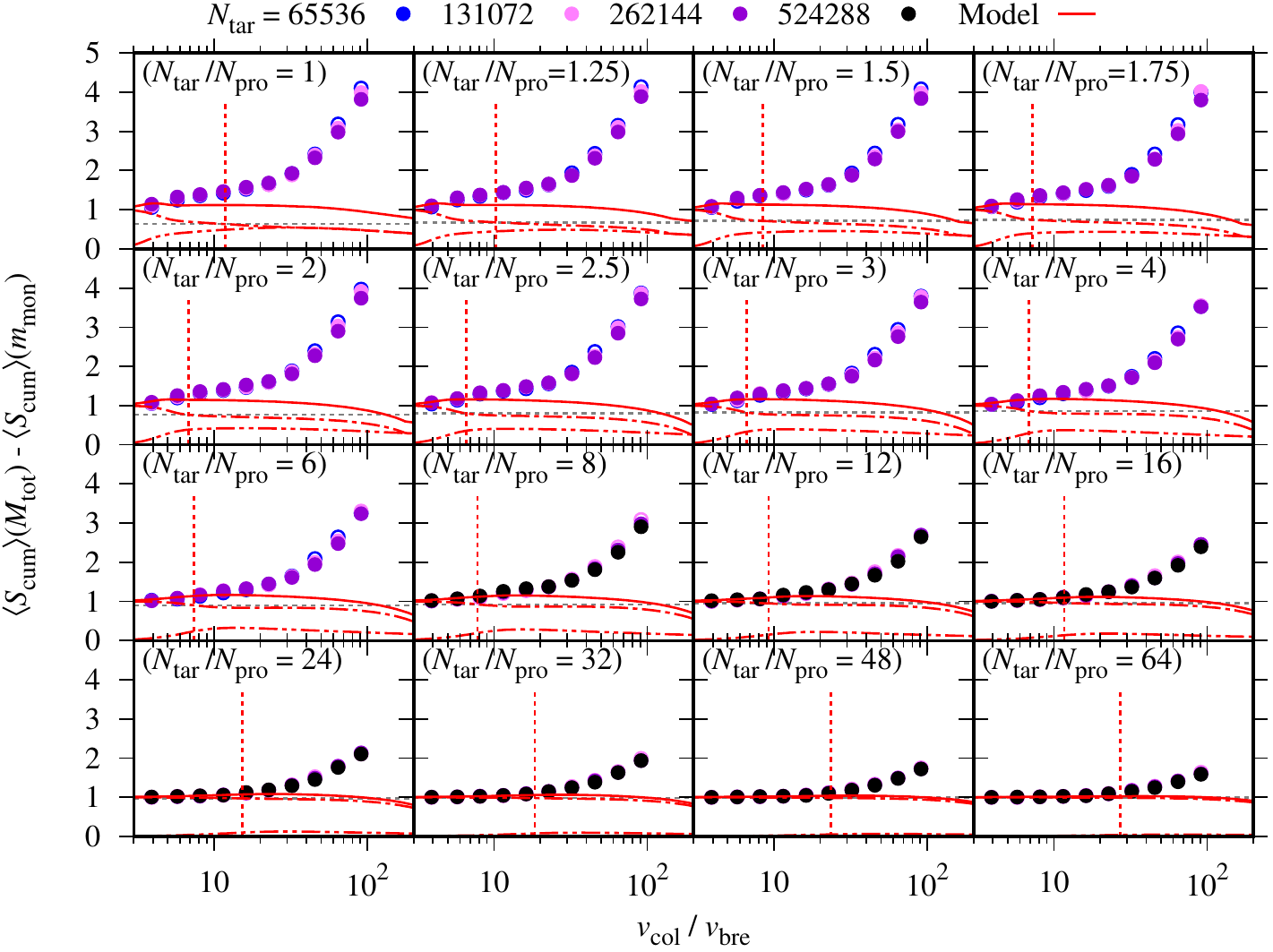}
  \caption{Ratio of the total cumulative geometric cross section of the fragments, except for monomer fragments, to the geometric cross section of a dust grain with $M = M_{\rm tot}$, $\langle S_{\rm cum} \rangle (M_{\rm tot}) - \langle S_{\rm cum} \rangle (m_{\rm mon})$, against the normalized collision velocity, $v_{\rm col} / v_{\rm bre}$ (circles). Colors except for gray and red are the same as Figure \ref{fig:fitting_vcob_N2ndoNpro_Ntar_Ntop_ylin_fit_model}. Horizontal gray dotted lines denote the geometric cross section of the target. Red dashed-dotted and chain double-dashed lines mark the geometric cross section of a dust grain with $M = \langle M_{\rm lar} \rangle$ and $\langle M_{\rm 2nd} \rangle$, obtained from our fragment model, respectively. Red solid line marks the sum of the geometric cross sections of the largest remnant and the second remnant. Vertical red dotted line marks $v_{\rm col} = v_{\rm fra}$ obtained from our fragment model. Different panels correspond to results of different initial mass ratios.}
  \label{fig:vcob_ScumNm1_Ntar_Ntop_ylin}
\end{figure}

Figure \ref{fig:vcob_ScumNm1_Ntar_Ntop_ylin} shows the ratio of the total cumulative geometric cross section of the fragments, except for monomer fragments, to the geometric cross section of a dust grain with $M = M_{\rm tot}$ (hereafter, called the total cross section ratio), $\langle S_{\rm cum} \rangle (M_{\rm tot}) - \langle S_{\rm cum} \rangle (m_{\rm mon})$, against the normalized collision velocity, $v_{\rm col} / v_{\rm bre}$, for different initial mass ratios.
The total cross section ratio roughly corresponds to $[S(\langle N_{\rm lar} \rangle) + S(\langle N_{\rm 2nd} \rangle)] / S(N_{\rm tot}) \approx 1$ for $v_{\rm col} < v_{\rm fra}$ and is about the same order of $S(\langle N_{\rm lar} \rangle) / S(N_{\rm tot}) \sim S(N_{\rm tar}) / S(N_{\rm tot})$ or $S(\langle N_{\rm 2nd} \rangle) / S(N_{\rm tot})$ for $v_{\rm col} \lesssim 30 v_{\rm bre} \approx 10^2 ~ \mathrm{m ~ s^{-1}}$.
For very high collision velocities, on the other hand, the total cross section ratio tends to increase with increasing $v_{\rm col}$.
This indicates that the geometric cross section is dominated by small fragments.
However, those results depend on $N_{\rm tar}$, and large dust aggregates tend to produce the small total cross section ratios.
For dust aggregates with sufficiently large masses to undergo collisional fragmentation in realistic protoplanetary disks, the geometric cross section would be dominated by relatively large fragments even for such a very high collision velocity.
Thus, our results suggest that the geometric cross section of the fragments after two-body collisions is dominated by relatively large fragments in realistic conditions of protoplanetary disks.

\subsection{Comparison with Previous Laboratory Experiments} \label{sec:dis_com_lab}

Takagi et al. (1984) carried out impact fragmentation experiments of basalts and pyrophyllites for $v_{\rm col} \sim 10^2$-$10^3 ~ \mathrm{m ~ s^{-1}}$.
They showed that the cumulative number distributions of the fragments not smaller than a given mass are divided into three regimes by the dependence of the slope on the mass, which are "regime I", "regime II", and "regime III" in order of decreasing mass of the fragments.
Mizutani et al. (1990) showed the same tendency for water ice, using data obtained by Lange \& Ahrens (1982).

\begin{figure}
  \plotone{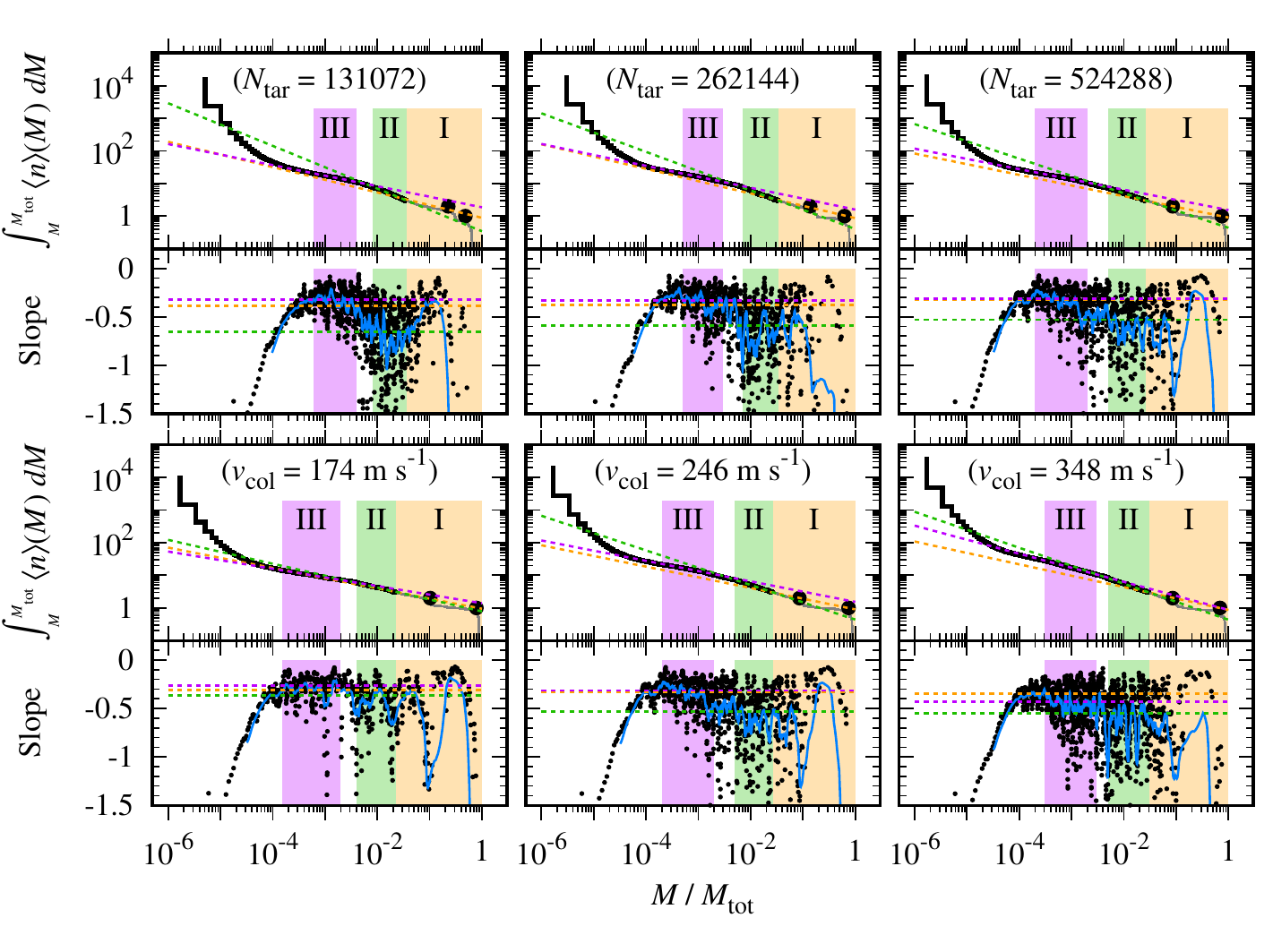}
  \caption{Cumulative number distribution of fragments not smaller than a given mass $M$, $\int _M^{M_{\rm tot}} \langle n \rangle (M) dM$ (upper row of each panel), and their slopes (lower row of each panel) for $N_{\rm tar} =$ 131072, 262144, and 524288 with $v_{\rm col} = 246 ~ \mathrm{m ~ s^{-1}}$ and $N_{\rm pro} = 65536$ (upper panels), and for $v_{\rm col} =$ 174, 246, and 348 $\mathrm{m ~ s^{-1}}$ with $N_{\rm tar} = 524288$ and $N_{\rm pro} = 65536$ (lower panels), respectively. The upper right panel is the same as the lower middle panel. Black circles in upper row of each panel mark $M = \langle M_{\rm lar} \rangle$ and $\langle M_{\rm 2nd} \rangle$. Black solid lines indicate the cumulative number distributions in regime II and regime III, and gray solid lines indicate the cumulative number distributions in regime I. Blue solid lines indicate averages of 32 data points. Other colors represent regime I (orange), regime II (green), and regime III (purple), respectively. Dotted lines indicate fits of the numerical results by power laws. Shaded regions represent ranges of data points employed in fits.}
  \label{fig:fitting_NoNt_slope_Ntar_vcol}
\end{figure}

Figure \ref{fig:fitting_NoNt_slope_Ntar_vcol} shows the cumulative number distributions of fragments not smaller than a given mass $M$, $\int _M^{M_{\rm tot}} \langle n \rangle (M) dM$, and their slopes, obtained from the results of our simulations.
Takagi et al. (1984) showed that the masses at inflection points of slopes decrease with increasing the mass ratio and increase with increasing the collision velocity, for the fixed projectile mass.
Our results imply the same tendency.

\begin{figure}
  \plotone{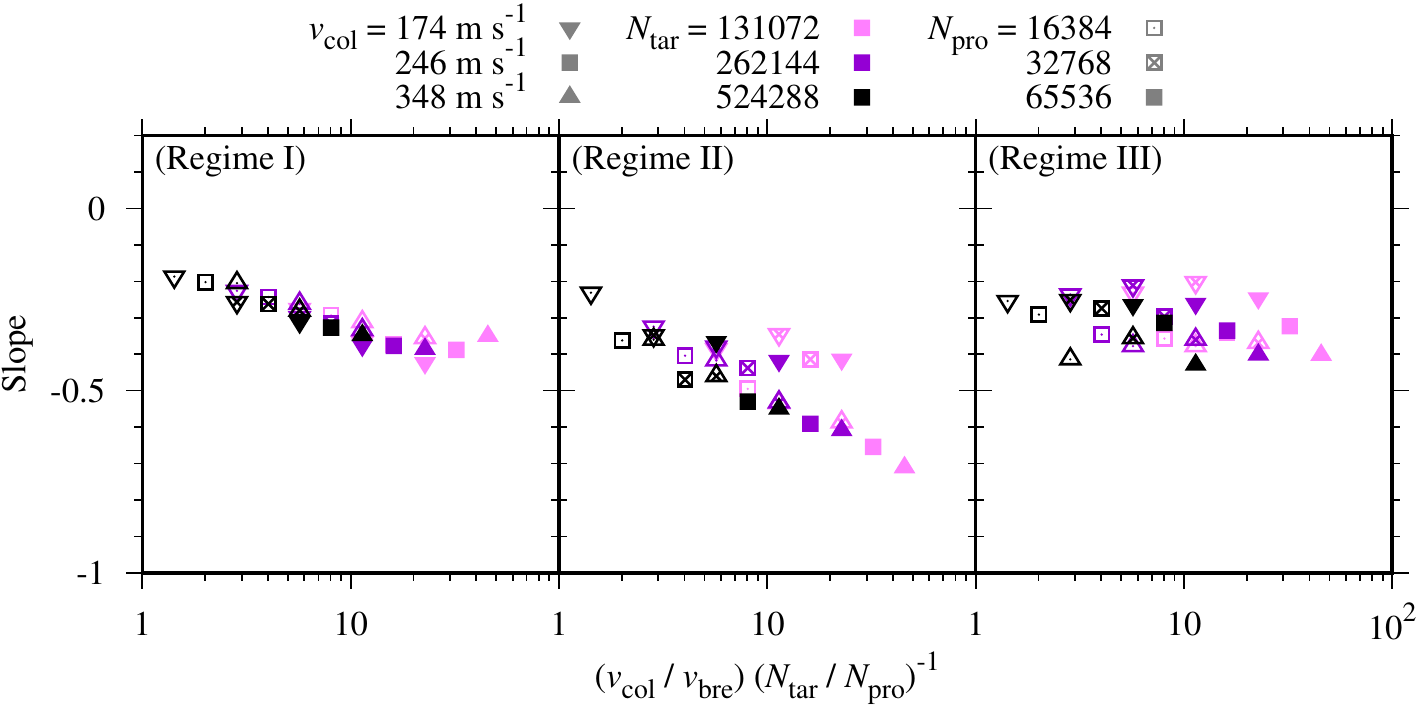}
  \caption{Slopes obtained from fits of the numerical results by power laws against $(v_{\rm col} / v_{\rm bre}) (N_{\rm tar} / N_{\rm pro})^{-1}$. Down-pointing triangles, squares, and up-pointing triangles indicate $v_{\rm col} =$ 174, 246, and 348 $\mathrm{m ~ s^{-1}}$, respectively. Colors represent $N_{\rm tar} =$ 131,072 (magenta), 262,144 (purple), and 524,288 (black), respectively. Open symbols, symbols with crosses, and filled symbols indicate $N_{\rm pro} =$ 16,384, 32,768, and 65,536, respectively. Different panels correspond to different regimes.}
  \label{fig:vcoboNtop_fittingslope_regime}
\end{figure}

Takagi et al. (1984) and Mizutani et al. (1990) showed that in regime I and regime II the slopes of the cumulative number distributions of fragments tend to decrease with increasing the non-dimensional impact stress, which increases with $v_{\rm col}$ and is inversely proportional to $N_{\rm tar} / N_{\rm pro}$.
On the other hand, the slope in regime III was almost constant against the non-dimensional impact stress except for the very high collision velocity of $> 6 \times 10^2 ~ \mathrm{m ~ s^{-1}}$ (Takagi et al. 1984).
Figure \ref{fig:vcoboNtop_fittingslope_regime} shows the slopes obtained from fits of the numerical results as a function of $(v_{\rm col} / v_{\rm bre}) (N_{\rm tar} / N_{\rm pro})^{-1}$.
While our results denote the same tendency as these previous studies, the absolute values of the slopes in regime I and regime III, which are $< 0.5$, are lower than those obtained by Takagi et al. (1984) and Mizutani et al. (1990), which are $\approx 1$ in regime I and $\approx 0.5$ in regime III.
The maximum collision velocity in this paper is 348 $\mathrm{m ~ s^{-1}}$ and is much lower than those of the previous studies, 990 $\mathrm{m ~ s^{-1}}$ for basalts and 950 $\mathrm{m ~ s^{-1}}$ for pyrophyllites (Takagi et al. 1984), and 1050 $\mathrm{m ~ s^{-1}}$ for ice (Lange \& Ahrens 1982).
Both the target and the projectile adopted in this paper are BPCA clusters and are different from ones used in Takagi et al. (1984) and Lange \& Ahrens (1982), e.g., very homogeneous natural pyrophyllites were used as targets in Takagi et al. (1984).
These may be the reasons for the difference between the results of our numerical simulations and the previous laboratory experiments.

Figure \ref{fig:vcoboNtop_fittingslope_regime} suggests that the slope in regime III is roughly constant against $(v_{\rm col} / v_{\rm bre}) (N_{\rm tar} / N_{\rm pro})^{-1}$ but decreases with increasing $v_{\rm col} / v_{\rm bre}$ and is independent of $N_{\rm tar} / N_{\rm pro}$.
This tendency to $v_{\rm col} / v_{\rm bre}$ and $N_{\rm tar} / N_{\rm pro}$ is the same as that of $B_{\rm cum}$ shown in Figure \ref{fig:fitting_Ntop_vcob_fittingABC_Ncum}(b).
On the other hand, Figure \ref{fig:vcoboNtop_fittingslope_regime} also suggests that the slope in regime II depends on $N_{\rm tar} / N_{\rm pro}$, and this is different from the dependence of $B_{\rm cum}$, which is primarily determined by the data points in the regime III.
For $v_{\rm col} \le 123 ~ \mathrm{m ~ s^{-1}}$, the slopes in regime III are seen to be roughly the same as those in regime II.
Takagi et al. (1984) showed that the mass at the inflection point from regime II to III continuously increases with increasing the collision velocity $v_{\rm col}$ for the fixed projectile mass.
On the other hand, the mass at the inflection point from regime I to II sensitively increases for the relatively low $v_{\rm col}$, whereas that hardly increases for the high $v_{\rm col}$.
Figure \ref{fig:fitting_NoNt_slope_Ntar_vcol} implies that our results show the same tendency as Takagi et al. (1984).
Then, regime II might disappear for the low collision velocity.
For $v_{\rm col} \ge 174 ~ \mathrm{m ~ s^{-1}}$, on the other hand, there is regime II at $M \sim 10^{-2} M_{\rm tot}$.
Thus, we employed data points with $M \le 3 \times 10^{-3} M_{\rm tot}$ in fits to derive $A_{\rm cum}$ and $B_{\rm cum}$ (Section \ref{sec:res_small}).
Figure \ref{fig:vcoboNtop_fittingslope_regime} shows that the slopes in regime II are clearly smaller than those in regime III for the low mass ratio and the high collision velocity.
This means that the cumulative mass distribution of the fragments, $\langle M_{\rm cum} \rangle (M)$, has a shallow slope around $\langle M_{\rm cum} \rangle (M) \sim \langle M_{\rm eje} \rangle$.
Thus, our analytic formula that neglects the contribution of regime II to the cumulative mass distribution of the fragments will underestimate the mass of the third-largest remnant for the high collision velocity of $v_{\rm col} > 10^2 ~ \mathrm{m ~ s^{-1}}$.

\subsection{Slope of Distribution of Fragments} \label{sec:dis_gra_dis}

So far we determined the fitting parameters with assuming the cumulative mass distribution of fragments, $\langle M_{\rm cum} \rangle (M)$ (Equation (\ref{equ:fitting_NcumoNtot})).
In order to check the consistency of this procedure, we take the opposite way of analysis in this section; we introduce a power-law profile of the number distribution of fragments (the integrand), $\langle n \rangle (M) = A' (M / m_{\rm mon})^{B - 2}$ and analyze the cumulative mass, geometric cross section, and number distributions of fragments mentioned in the above sections.
We note that Arakawa et al. (2022) also performed a similar analyses (with $\langle n \rangle (N) \propto N^{- \alpha}$).

\subsubsection{Cumulative Mass Distribution}

For $B \neq 0$, $\langle M_{\rm cum} \rangle = \int _{m_{\rm mon}}^M M \langle n \rangle (M) dM = (A' m_{\rm mon}^2 / B) (N^B - 1)$.
In the case with $N \gg 1$ and $B \neq 0$, $\langle M_{\rm cum} \rangle \approx (A' m_{\rm mon}^2 / B) N^B$ for $B > 0$ or $\approx A' m_{\rm mon}^2 / (-B)$ for $B < 0$.
Thus, most of the total mass is dominated by the masses of large remnants for $B > 0$ or by small fragments for $B < 0$.

We neglected the detailed cumulative mass profile of tiny fragments and assumed the fitting formula (Equation (\ref{equ:fitting_NcumoNtot})) in Section \ref{sec:res_small}.
However, Figure \ref{fig:fitting_NoNt_cumlative_Ntar_v5_Ntop8}(a) indicates that the slope of $M^2 \langle n \rangle (M)$ is negative for $N \lesssim 10^2$.
We assume the integrand in the cumulative number distribution of fragments to be $\langle n \rangle (M) = A'_{\rm L} (M / m_{\rm mon})^{-B_{\rm L} - 2} + A'_{\rm H} (M / m_{\rm mon})^{B_{\rm H} - 2}$, where $B_{\rm L} > 0$ and $B_{\rm H} > 0$.
Then the cumulative mass distribution of fragments $\langle M_{\rm cum} \rangle$ is estimated to be $\approx (A'_{\rm H} m_{\rm mon}^2 / B_{\rm H}) N^{B_{\rm H}} + A'_{\rm L} m_{\rm mon}^2 / B_{\rm L}$ at $N \gg 1$ and has the same formula as Equation (\ref{equ:fitting_NcumoNtot}), $\langle M_{\rm cum} \rangle = A_{\rm cum} M_{\rm tot} (M / M_{\rm tot})^{B_{\rm cum}} + C_{\rm cum} M_{\rm tot} = (A_{\rm cum} m_{\rm mon} / N_{\rm tot}^{B_{\rm cum} - 1}) N^{B_{\rm cum}} + C_{\rm cum} N_{\rm tot} m_{\rm mon}$.
Thus, we have $A_{\rm cum} = A'_{\rm H} N_{\rm tot}^{B_{\rm cum} - 1} m_{\rm mon} / B_{\rm H}$, $B_{\rm cum} = B_{\rm H}$, and $C_{\rm cum} = A'_{\rm L} m_{\rm mon} / B_{\rm L} N_{\rm tot}$.

From Figure \ref{fig:fitting_NoNt_cumlative_Ntar_v5_Ntop8}(a), we also have $M^2 \langle n \rangle (M) / M_{\rm tot} = (A'_{\rm L} m_{\rm mon} / N_{\rm tot}^{B_{\rm L} + 1}) (M / M_{\rm tot})^{-B_{\rm L}} + A'_{\rm H} N_{\rm tot}^{B_{\rm H} - 1} m_{\rm mon} (M / M_{\rm tot})^{B_{\rm H}}$.
Figure \ref{fig:fitting_NoNt_cumlative_Ntar_v5_Ntop8}(a) suggests that $A'_{\rm H} N_{\rm tot}^{B_{\rm H} - 1} m_{\rm mon}$ and $B_{\rm H}$ are independent of $N_{\rm tot}$.
These tendencies show that $A_{\rm cum} = A'_{\rm H} N_{\rm tot}^{B_{\rm H} - 1} m_{\rm mon} / B_{\rm H}$ and $B_{\rm cum} = B_{\rm H}$ are also independent of $N_{\rm tot}$.
Figure \ref{fig:fitting_NoNt_cumlative_Ntar_v5_Ntop8}(a) and our other results also indicate that $A'_{\rm L} m_{\rm mon} / N_{\rm tot}^{B_{\rm L} + 1}$ tends to decrease with increasing $N_{\rm tot}$ and the dependence of that on $N_{\rm tot}$ is seen to be less than $-1$, and that $B_{\rm L}$ is seen to be independent of $N_{\rm tot}$ and $\lesssim 1$.
These tendencies imply that $C_{\rm cum} = (A'_{\rm L} m_{\rm mon} / N_{\rm tot}^{B_{\rm L} + 1}) N_{\rm tot}^{B_{\rm L}} / B_{\rm L}$ decreases with increasing $N_{\rm tot}$.
This dependence of $C_{\rm cum}$ on $N_{\rm tot}$ is the same as one shown in Figure \ref{fig:fitting_Ntop_vcob_fittingABC_Ncum}(d), $C_{\rm cum} \propto N_{\rm tot}^{Q_{\rm Ctot}}$ with $Q_{\rm Ctot} < 0$.
This tendency and $B_{\rm H} = B_{\rm cum} > 0$ as shown in Figure \ref{fig:fitting_Ntop_vcob_fittingABC_Ncum}(b) suggest that the masses of large remnants dominate most of the total mass for collisions between enough large dust aggregates to undergo collisional fragmentation in realistic protoplanetary disks.

\subsubsection{Geometric Cross section}

The ratio of geometric cross sections of fragments derived from Equation (\ref{equ:Scumulative}) in Section \ref{sec:dis_cro_sec} is given by $\langle S_{\rm cum} \rangle = \int _{m_{\rm mon}}^M (M / M_{\rm tot})^{2/3} \langle n \rangle (M) dM \approx [A'_{\rm H} N_{\rm tot}^{B_{\rm H} - 1} m_{\rm mon} / (B_{\rm H} - 1/3)] (M / M_{\rm tot})^{B_{\rm H} - 1/3}$ at $N \gg 1$ for $N_{\rm tot} \rightarrow \infty$.
This shows that the power-law component of $\langle S_{\rm cum} \rangle$ is independent of $N_{\rm tot}$ because $A'_{\rm H} N_{\rm tot}^{B_{\rm H} - 1} m_{\rm mon}$ and $B_{\rm H}$ are independent of $N_{\rm tot}$.
The condition, when the total geometric cross section of the fragments is dominated by the large remnants, is $B_{\rm H} - 1/3 > 0$, i.e., $B_{\rm H} = B_{\rm cum} > 1/3$.
Since Figure \ref{fig:fitting_Ntop_vcob_fittingABC_Ncum}(b) indicates $B_{\rm cum} > 1/3$ for $v_{\rm col} < 4 \times 10^2 ~ \mathrm{m ~ s^{-1}}$, the large remnants will dominate the cumulative geometric cross section of all fragments for collisions of dust aggregates in protoplanetary disks.
For collisions without perfect sticking, Figures \ref{fig:fitting_vcob_N2ndoNpro_Ntar_Ntop_ylin_fit_model} and \ref{fig:vcob_ScumNm1_Ntar_Ntop_ylin} show $\langle M_{\rm lar} \rangle \sim M_{\rm tar}$ and $\langle M_{\rm 2nd} \rangle \lesssim M_{\rm pro}$ for $N_{\rm tar} / N_{\rm pro} \sim 1$ or $\langle M_{\rm 2nd} \rangle \ll M_{\rm pro}$ for $N_{\rm tar} / N_{\rm pro} \gtrsim 10^1$ for $v_{\rm col} \lesssim v_{\rm fra}$, and $\langle M_{\rm lar} \rangle \sim M_{\rm pro}$ for $N_{\rm tar} / N_{\rm pro} \sim 1$ or $\langle M_{\rm lar} \rangle \sim M_{\rm tar}$ for $N_{\rm tar} / N_{\rm pro} \gtrsim 10^1$ and $\langle M_{\rm 2nd} \rangle \sim M_{\rm pro}$ for $v_{\rm col} \gtrsim v_{\rm fra}$.
This order estimation of the masses suggests that the total geometric cross section of the fragments will be then estimated to be about the order of the geometric cross section of the target or the projectile.
For the much higher collision velocity than one in this paper, on the other hand, $\langle M_{\rm lar} \rangle$ and $\langle M_{\rm 2nd} \rangle$ will become much smaller than $M_{\rm tar}$ and $M_{\rm pro}$, respectively.
However, dust monomers will be fragmented and/or distorted in such a case.

\subsubsection{Number Distribution}

The cumulative number distribution of fragments not smaller than a given mass $M$, shown in Section \ref{sec:dis_com_lab}, is estimated to be $\int _M^{M_{\rm tot}} \langle n \rangle (M) dM \approx [A'_{\rm L} m_{\rm mon} / (1 + B_{\rm L})] N^{-(1 + B_{\rm L})} + \int _M^{M_{\rm tot}} A'_{\rm H} (M / m_{\rm mon})^{B_{\rm H} - 2} dM$.
For $v_{\rm col} > 10^2 ~ \mathrm{m ~ s^{-1}}$ (Figure \ref{fig:fitting_NoNt_slope_Ntar_vcol}), $\int _M^{M_{\rm tot}} A'_{\rm H} (M / m_{\rm mon})^{B_{\rm H} - 2} dM \approx [A'_{\rm H} m_{\rm mon} / (1 - B_{\rm H})] N^{-(1 - B_{\rm H})}$ since Figure \ref{fig:fitting_Ntop_vcob_fittingABC_Ncum}(b) estimates $B_{\rm cum} = B_{\rm H} < 1$ at such a high collision velocity, and the cumulative number distribution of the fragments is dominated by the number of the small fragments even for fragments except for tiny ones, $1 \ll N \ll N_{\rm tot}$.
On the other hand, $\int _M^{M_{\rm tot}} A'_{\rm H} (M / m_{\rm mon})^{B_{\rm H} - 2} dM \approx [A'_{\rm H} m_{\rm mon} / (B_{\rm H} - 1)] N_{\rm tot}^{B_{\rm H} - 1} = A_{\rm cum} B_{\rm cum} / (B_{\rm cum} - 1) =$ const for $v_{\rm col} \ll 10^2 ~ \mathrm{m ~ s^{-1}}$ because of $B_{\rm cum} = B_{\rm H} > 1$.
The collision velocity with $B_{\rm cum} = B_{\rm H} = 1$ is calculated by Equation (\ref{equ:fitting_Bcum}), and is given by $v_{\rm col} = 56 ~ \mathrm{m ~ s^{-1}}$ and is about $v_{\rm fra} = 52 ~ \mathrm{m ~ s^{-1}}$ for $N_{\rm tar} / N_{\rm pro} = 1$.

\subsection{Material Properties of Water Ice} \label{sec:dis_mat_ice}

The material properties of water ice adopted in this paper are the same as those adopted in previous studies (Wada et al. 2007, 2009, 2013; Paper I; see also Section \ref{sec:pro}).
The surface energy of water ice adopted in this paper is 100 $\mathrm{mJ ~ m^{-2}}$ (Table \ref{tab:properties}), which corresponds to one at relatively high temperatures of $\approx$ 200 K (Musiolik \& Wurm 2019).
Musiolik \& Wurm (2019) showed that the surface energy of water ice rapidly decreases with decreasing the temperature of $\lesssim$ 200 K.
Their results suggest that water-ice dust aggregates in the colder outer region of protoplanetary disks are more fragile than those in the hotter inner region.
In order to examine collisional outcomes of relatively cold icy dust aggregates, we carry out numerical simulations of collisions between two dust aggregates composed of dust monomers with lower surface energy of 50 $\mathrm{mJ ~ m^{-2}}$.
In this case, the characteristic break-up velocity is given by $v_{\rm bre} = 2.1 ~ \mathrm{m ~ s^{-1}}$.

\begin{figure}
  \plotone{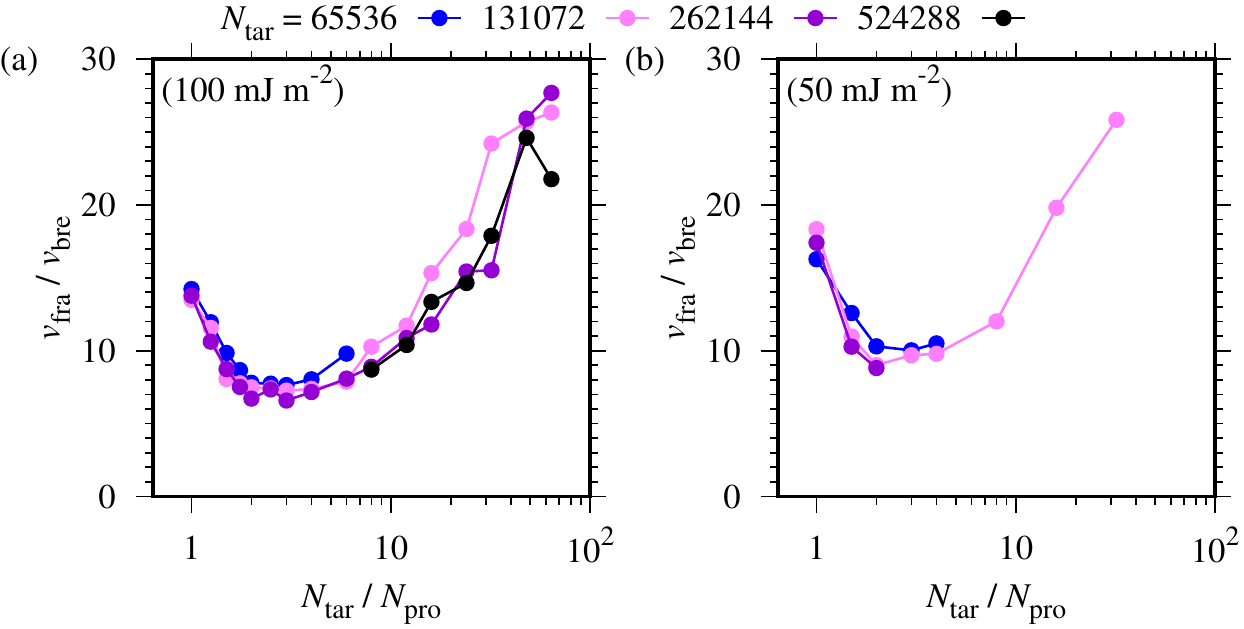}
  \caption{Critical collisional fragmentation velocity normalized by the characteristic break-up velocity, $v_{\rm fra} / v_{\rm bre}$, for the surface energy of 100 $\mathrm{mJ ~ m^{-2}}$ (a) and 50 $\mathrm{mJ ~ m^{-2}}$ (b) against the mass ratio before the collision, $N_{\rm tar} / N_{\rm pro}$ (filled circles and solid lines). Colors are the same as in Figure \ref{fig:fitting_vcob_N2ndoNpro_Ntar_Ntop_ylin_fit_model}.}
  \label{fig:Ntop_vfob_Ntar_iag5}
\end{figure}

Figure \ref{fig:Ntop_vfob_Ntar_iag5} compares the critical collisional fragmentation velocities $v_{\rm fra}$ of hot (100 $\mathrm{mJ ~ m^{-2}}$) and cold (50 $\mathrm{mJ ~ m^{-2}}$) icy dust aggregates.
The results with the low surface energy of 50 $\mathrm{mJ ~ m^{-2}}$ (Figure \ref{fig:Ntop_vfob_Ntar_iag5}(b)) illustrate the same tendency as that with 100 $\mathrm{mJ ~ m^{-2}}$ (Figure \ref{fig:Ntop_vfob_Ntar_iag5}(a) and Paper I).
In both panels of Figure \ref{fig:Ntop_vfob_Ntar_iag5}, the mass transfer from the target to the projectile occurs in unequal-mass collisions, and the critical collisional fragmentation velocity $v_{\rm fra}$ with the mass ratio of $\approx$ 3 is about half of that obtained from equal-mass collisions.
In addition, the critical collisional fragmentation velocities for the surface energy of 50 $\mathrm{mJ ~ m^{-2}}$ is lower than those for 100 $\mathrm{mJ ~ m^{-2}}$ since dust aggregates with the lower surface energy is more fragile; $v_{\rm bre} = 2.1 ~ \mathrm{m ~ s^{-1}}$ for 50 $\mathrm{mJ ~ m^{-2}}$ is lower than $v_{\rm bre} = 3.8 ~ \mathrm{m ~ s^{-1}}$ for 100 $\mathrm{mJ ~ m^{-2}}$.
On the other hand, the critical collisional fragmentation velocities normalized by the characteristic break-up velocity for 50 $\mathrm{mJ ~ m^{-2}}$ are seen to be slightly higher than those for 100 $\mathrm{mJ ~ m^{-2}}$.
However, it is still uncertain whether this tendency of the increasing normalized critical velocity with decreasing surface energy continues to further lower surface energy or not.
We plan to perform simulations of collisions between much more fragile dust aggregates in future studies.

\section{Summary} \label{sec:sum}

Collisional fragmentation of dust aggregates prevents or hinders the dust growth through collisional sticking in the process of the planet formation.
Not to mention the case when the dust aggregates are prevented from growing, it may be also a serious issue for the planet formation even if the dust growth is only delayed.
Since dust aggregates generally drift toward the central star in the protoplanetary disk (Adachi et al. 1976; Weidenschilling 1977; Brauer et al. 2008; Birnstiel et al. 2010; Okuzumi et al. 2012), the planet will not be formed if dust aggregates fall toward the central star before the growth of the dust aggregates to planetesimals owing to the delay in the dust growth.
The outcome of the collision between two dust aggregates depends on the collision velocity, the mass ratio, the material properties, the internal structure, etc. of colliding dust aggregates.
In order to clarify whether or not dust aggregates can grow via their collisional sticking in the protoplanetary disk, we should know the detailed output of collisional outcomes against those input parameters and utilize it for studies of the dust evolution.

In this paper, we carried out numerical $N$-body simulations of collisions between water-ice BPCA clusters and modeled the mass distribution of the fragments to be among the collisional outcomes as functions of the collision velocity, the mass ratio, and the total mass of the colliding bodies, which can be applied to the statistical simulations of the dust evolution, with using fits of the numerical results.
Our results are summarized as follows.

\begin{enumerate}
  \item
In our fragment model, we divide the fragments into three classes, which are the largest remnant, the second-largest remnant, and other ejecta except for these two large remnants (see Figure \ref{fig:diagram}).
  \item
The analytic formulae of the second remnant mass and the total other ejecta mass are given by the combination of the power-law functions of the collision velocity (Equations (\ref{equ:fitting_N2ndoNpro_model}) and (\ref{equ:fitting_E1a2oNpro_model})) using coefficients and power-law indices to be the functions of the mass ratio (Equations (\ref{equ:fitting_ASLandBSL_model}) and (\ref{equ:fitting_AELandBEL_model})).
Both analytic formulae consist of three velocity components, while the middle-velocity component little contributes to the analytic formulae at the mass ratio of $\ge 6$ for the second remnant mass and $\ge 12$ for the total other ejecta mass, respectively.
The mass of the largest remnant is derived from those analytic formulae and the total mass of the colliding bodies.
Our fragment model can roughly reproduce the masses of the largest remnant and the second remnant obtained by our simulations.
  \item
Other ejecta are classified into fragments composed of multiple dust monomers and monomer fragments.
The cumulative mass distribution of fragments is fitted by the sum of the power-law mass distribution of fragments and the intercept corresponding to monomer fragments (Equation (\ref{equ:fitting_NcumoNtot})).
The power-law index of the cumulative mass distribution of fragments is dependent on the collision velocity only and independent of mass transfer from the target to the projectile.
  \item
The mass fraction of monomer fragments gradually decreases with increasing the total mass of the colliding bodies for a fixed mass ratio.
This tendency implies that multiple hierarchical disruptive collisions (i.e., collisions between fragments, collisions between fragments of fragments) are required for creating a great amount of individual dust monomers via collisional disruptions.
Applying our fragment model to the analysis of the geometric cross sections of the fragments ejected by a collision between dust aggregates in the protoplanetary disk, it is suggested that the total geometric cross section of the fragments is estimated to be comparable to the geometric cross section of the target.
\end{enumerate}

In this paper, we adopted submicron-sized icy dust monomers only.
Dust monomers in realistic protoplanetary disks are made of not only water ice but also silicate, $\mathrm{CO}_2$ ice, CO ice, organic matter, and so on (Kouchi et al. 2002, 2021; Kudo et al. 2002; Musiolik et al. 2016a, 2016b; Gail \& Trieloff 2017; Piani et al. 2017; Pinilla et al. 2017).
It is important to study the collisional outcomes of dust aggregates composed of dust monomers made of materials other than water ice.
We also adopted only BPCA clusters as initial colliding bodies in this paper.
The structure of the initial colliding bodies, e.g., the bulk densities and the coordination number, affects the collisional outcomes of dust aggregates (Wada et al. 2009, 2011; Seizinger et al. 2013).
We will address those topics in future works.






\acknowledgments

We thank Sota Arakawa for fruitful discussions.
This work was supported by JSPS KAKENHI Grant Nos. 21H00033, 22H01263 (T. K. Suzuki), 19K03941 (H. Tanaka), 17H01103, 17K05632, 18H05438, 18H05436, 20H04612, 21K03642, 22H00179, 22H01278 (H. Kobayashi) and 17H01105 (T. K. S. and H. K.), and by Program for Promoting Research on the Supercomputer Fugaku by the RIKEN Center for Computational Science (Toward a unified view of the universe: from large-scale structures to planets, grant 20351188 -- PI J. Makino) from the MEXT of Japan (T. K. S.).
Most of our numerical computations were carried out on the general-purpose PC cluster at Center for Computational Astrophysics, National Astronomical Observatory of Japan.

%






\appendix

\section{Specific Fitting Procedure} \label{sec:app_spe_pro}

We describe the specific procedures for the numerical fitting to derive $\langle M_{\rm 2nd} \rangle$ in Section \ref{sec:res_large}.

First, we fit the normalized second remnant mass $\langle M_{\rm 2nd} \rangle / M_{\rm pro}$ by Equation (\ref{equ:fitting_N2ndoNpro}) and obtain the fitting parameters $A_{\rm SL}$, $A_{\rm SM}$, $A_{\rm SH}$, $B_{\rm SL}$, $B_{\rm SM}$, and $B_{\rm SH}$.
Second, we fit the power-law index of the high-velocity component, $B_{\rm SH}$, by an assumption of a constant form, $B_{\rm SH} =$ const, and obtain the constant value $B_{\rm SH}$.
Third, we re-fit $\langle M_{\rm 2nd} \rangle / M_{\rm pro}$ by Equation (\ref{equ:fitting_N2ndoNpro}) with using constant $B_{\rm SH} = -0.529$ and obtain the re-fitting parameters $A_{\rm SL}$, $A_{\rm SM}$, $A_{\rm SH}$, $B_{\rm SL}$, and $B_{\rm SM}$.
Finally, as shown in Section \ref{sec:res_large}, we fit the re-fitting parameters and construct the analytic formula of $\langle M_{\rm 2nd} \rangle / M_{\rm pro}$, given by Equation (\ref{equ:fitting_N2ndoNpro_model}), from re-fitting parameters $A_{\rm SL}$, $A_{\rm SM}$, $A_{\rm SH}$, $B_{\rm SL}$, and $B_{\rm SM}$, and a constant fitting parameter $B_{\rm SH}$.

\begin{figure}
  \plotone{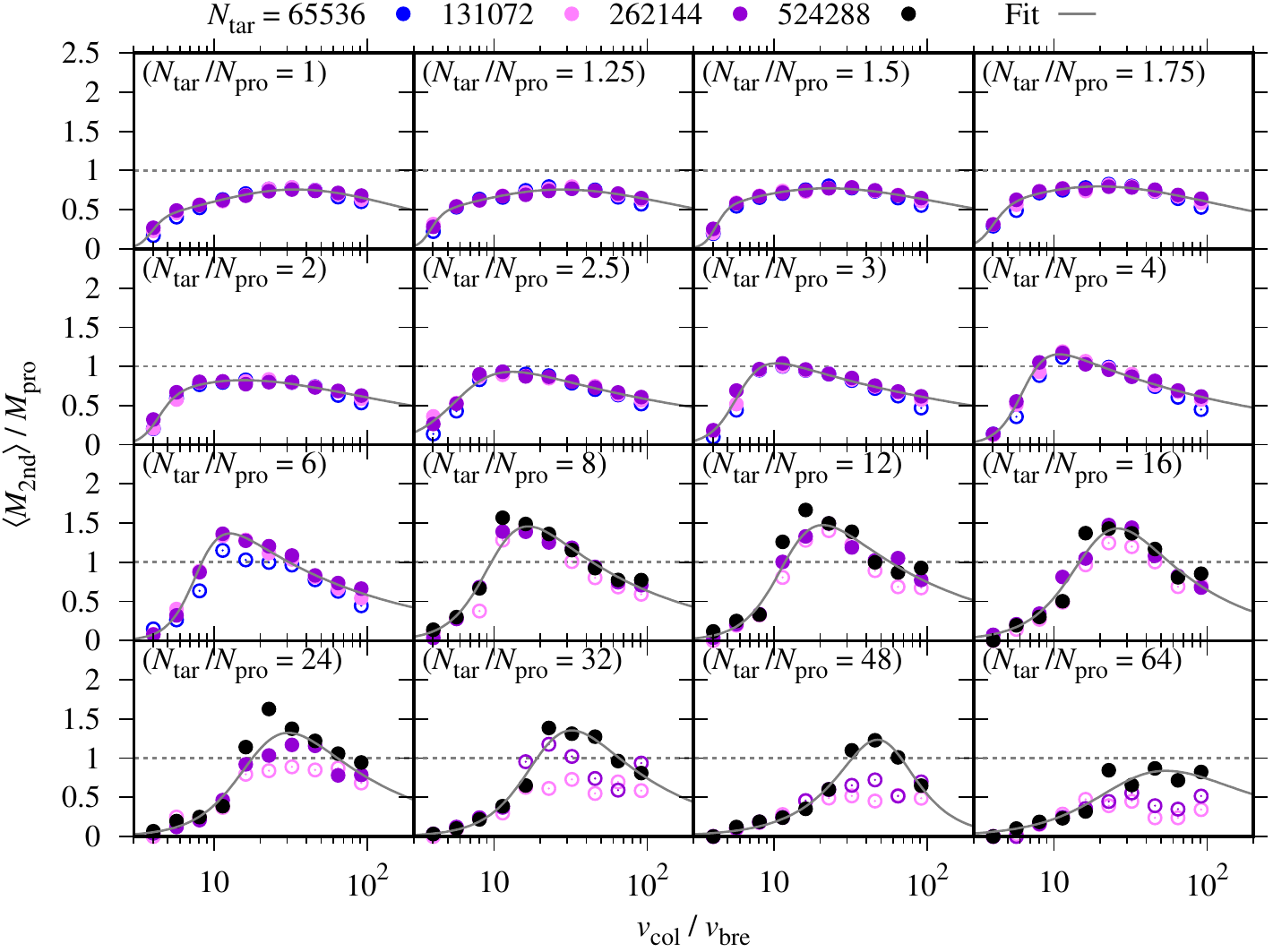}
  \caption{Same as Figure \ref{fig:fitting_vcob_N2ndoNpro_Ntar_Ntop_ylin_fit_model}, but $B_{\rm SH}$ is not assumed to be constant but a fitting parameter.}
  \label{fig:appendix_fitting_vcob_N2ndoNpro_Ntar_Ntop_ylin_dir}
\end{figure}

\begin{figure}
  \plotone{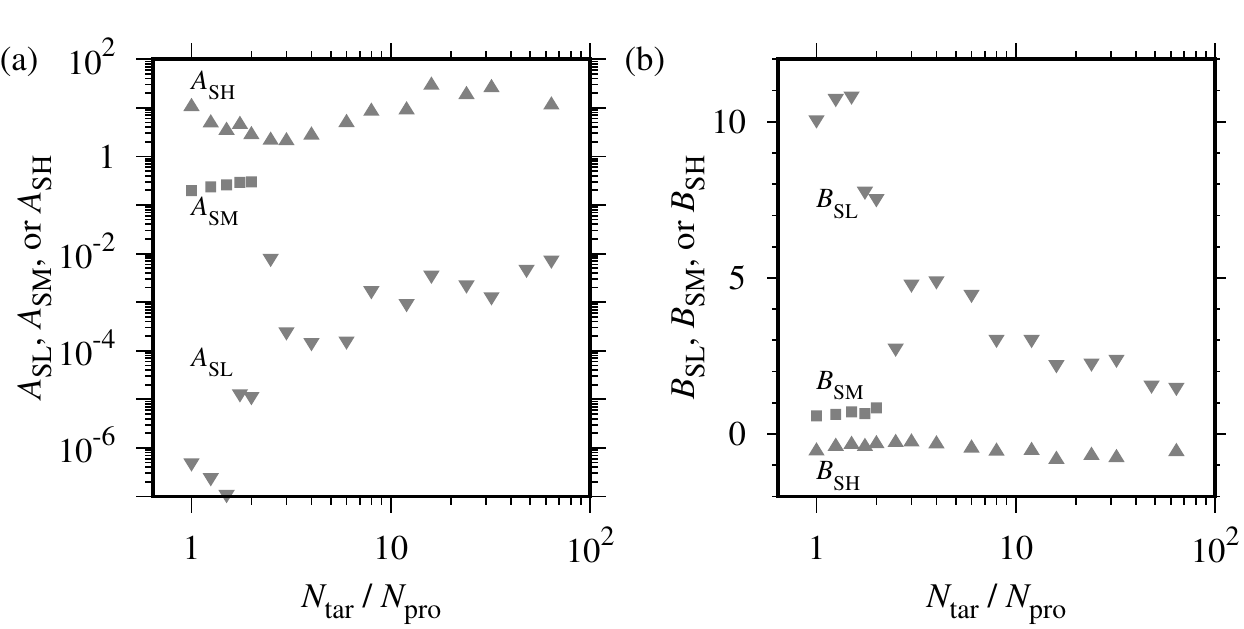}
  \caption{Same as Figure \ref{fig:fitting_Ntop_fittingparameters_N2nd_model}, but for fits without constant $B_{\rm SH}$.}
  \label{fig:appendix_fitting_Ntop_fittingparameters_N2nd}
\end{figure}

Figure \ref{fig:appendix_fitting_vcob_N2ndoNpro_Ntar_Ntop_ylin_dir} shows the normalized second remnant mass $\langle M_{\rm 2nd} \rangle / M_{\rm pro}$ against the normalized collision velocity $v_{\rm col} / v_{\rm bre}$ for different initial mass ratios $N_{\rm tar} / N_{\rm pro}$.
We confirm that the second remnant mass $\langle M_{\rm 2nd} \rangle$ contains the shallow power-law function given by $A_{\rm SM}$ and $B_{\rm SM}$ for middle velocities for $N_{\rm tar} / N_{\rm pro} \le 2$ and does not for $N_{\rm tar} / N_{\rm pro} \ge 2.5$.
Figure \ref{fig:appendix_fitting_Ntop_fittingparameters_N2nd} shows coefficients and power-law indices of Equation (\ref{equ:fitting_N2ndoNpro}) for fits without assuming constant $B_{\rm SH}$ as functions of the initial mass ratio.
The values of $B_{\rm SH}$ in Figure \ref{fig:appendix_fitting_Ntop_fittingparameters_N2nd} are the same as ones in Figure \ref{fig:fitting_Ntop_fittingparameters_N2nd_model}.

Next, we describe the specific fitting procedure to drive $\langle M_{\rm eje} \rangle$ in Section \ref{sec:res_large}.

First, we fit the normalized total other ejecta mass $\langle M_{\rm eje} \rangle / M_{\rm pro}$ by Equation (\ref{equ:fitting_E1a2oNpro}) and obtain the fitting parameters $A_{\rm EL}$, $A_{\rm EM}$, $A_{\rm EH}$, $B_{\rm EL}$, $B_{\rm EM}$, and $B_{\rm EH}$.
Second, we fit the power-law index of the high-velocity component, $B_{\rm EH}$, by an assumption of a simple power-law form and obtain $B_{\rm EH} = 1.12 (N_{\rm tar} / N_{\rm pro})^{0.0834}$.
Third, we re-fit $\langle M_{\rm eje} \rangle / M_{\rm pro}$ by Equation (\ref{equ:fitting_E1a2oNpro}) with using power-law $B_{\rm EH} = 1.12 (N_{\rm tar} / N_{\rm pro})^{0.0834}$ and obtain the re-fitting parameters $A_{\rm EL}$, $A_{\rm EM}$, $A_{\rm EH}$, $B_{\rm EL}$, and $B_{\rm EM}$.
Finally, as shown in Section \ref{sec:res_large}, we fit the re-fitting parameters and construct the analytic formula of $\langle M_{\rm eje} \rangle / M_{\rm pro}$, given by Equation (\ref{equ:fitting_E1a2oNpro_model}), from re-fitting parameters $A_{\rm EL}$, $A_{\rm EM}$, $A_{\rm EH}$, $B_{\rm EL}$, and $B_{\rm EM}$, and a power-law fitting parameter $B_{\rm EH}$.

\begin{figure}
  \plotone{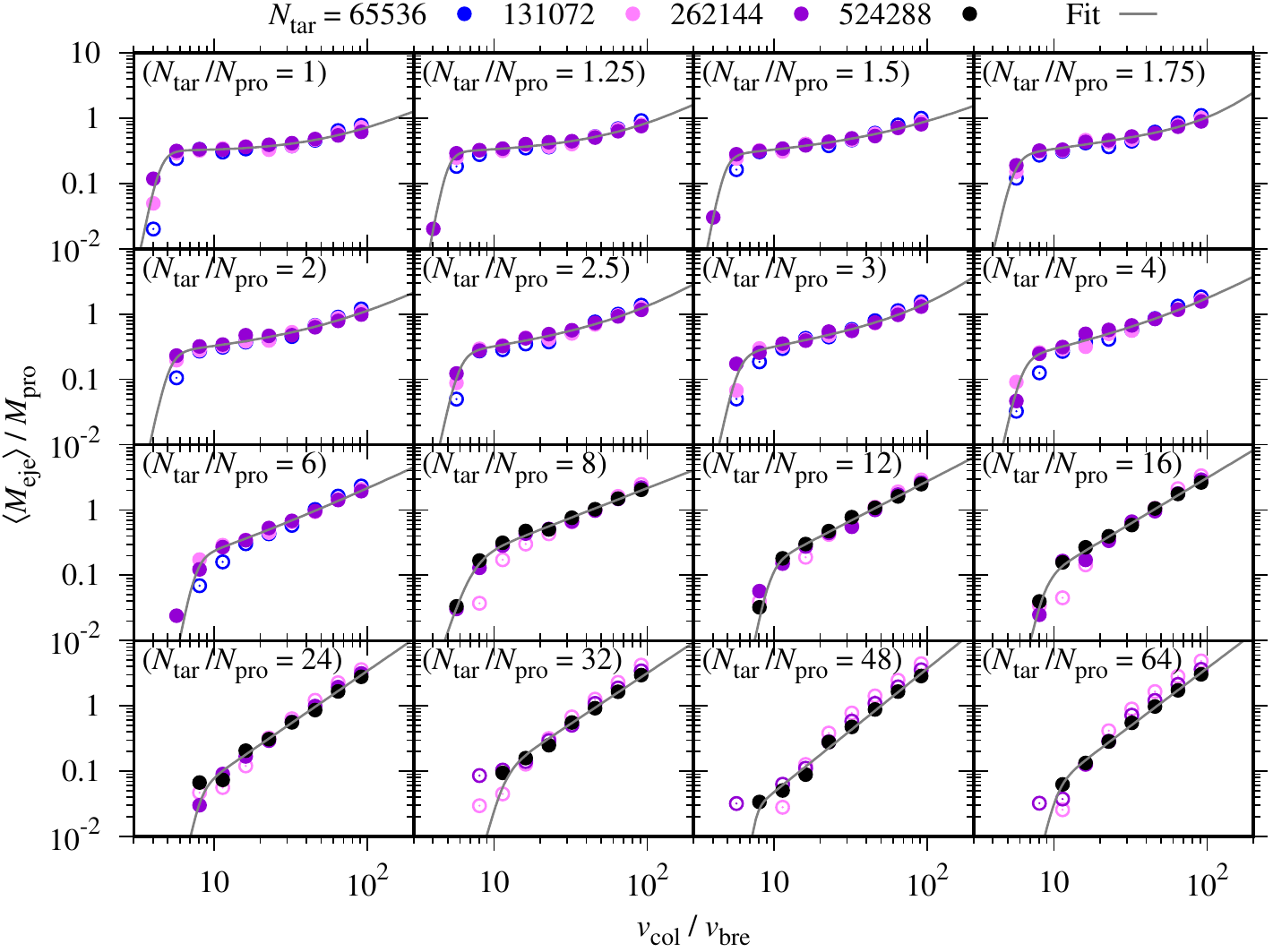}
  \caption{Same as Figure \ref{fig:fitting_vcob_E1a2oNpro_Ntar_Ntop_ylog_fit_model}, but $B_{\rm EH}$ is not assumed to be given by a simple power law but a fitting parameter.}
  \label{fig:appendix_fitting_vcob_E1a2oNpro_Ntar_Ntop_ylog_dir}
\end{figure}

\begin{figure}
  \plotone{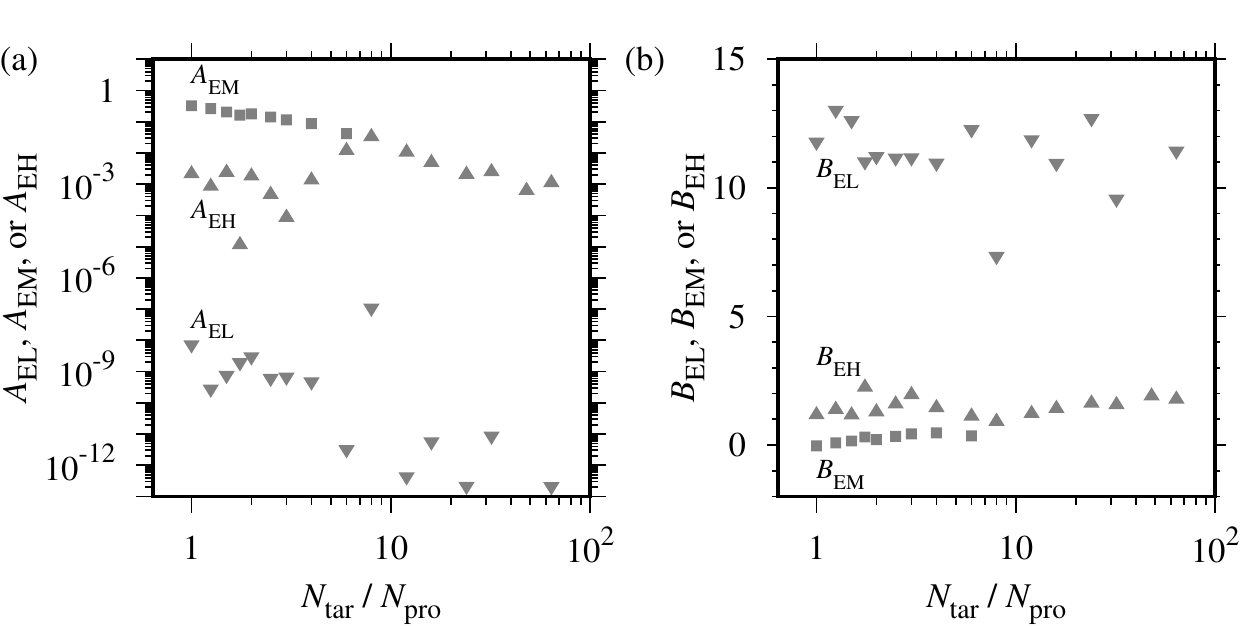}
  \caption{Same as Figure \ref{fig:fitting_Ntop_fittingparameters_E1a2_model}, but for fits without an assumption of a simple power-law $B_{\rm EH}$.}
  \label{fig:appendix_fitting_Ntop_fittingparameters_E1a2}
\end{figure}

Figure \ref{fig:appendix_fitting_vcob_E1a2oNpro_Ntar_Ntop_ylog_dir} shows the normalized total other ejecta mass $\langle M_{\rm eje} \rangle / M_{\rm pro}$ against the normalized collision velocity $v_{\rm col} / v_{\rm bre}$ for different initial mass ratios $N_{\rm tar} / N_{\rm pro}$.
We confirm that the total other ejecta mass $\langle M_{\rm eje} \rangle$ contains the shallow power-law function given by $A_{\rm EM}$ and $B_{\rm EM}$ for middle velocities for $N_{\rm tar} / N_{\rm pro} \le 6$ and does not for $N_{\rm tar} / N_{\rm pro} \ge 8$.
Figure \ref{fig:appendix_fitting_Ntop_fittingparameters_E1a2} shows coefficients and power-law indices of Equation (\ref{equ:fitting_E1a2oNpro}) for fits without assuming the simple power-law $B_{\rm EH}$ as functions of the initial mass ratio.
The values of $B_{\rm EH}$ in Figure \ref{fig:appendix_fitting_Ntop_fittingparameters_E1a2} are the same as ones in Figure \ref{fig:fitting_Ntop_fittingparameters_E1a2_model}.

\section{Reproduction of Numerical Results by Our Fragment Model} \label{sec:app_res}

\begin{figure}
  \plotone{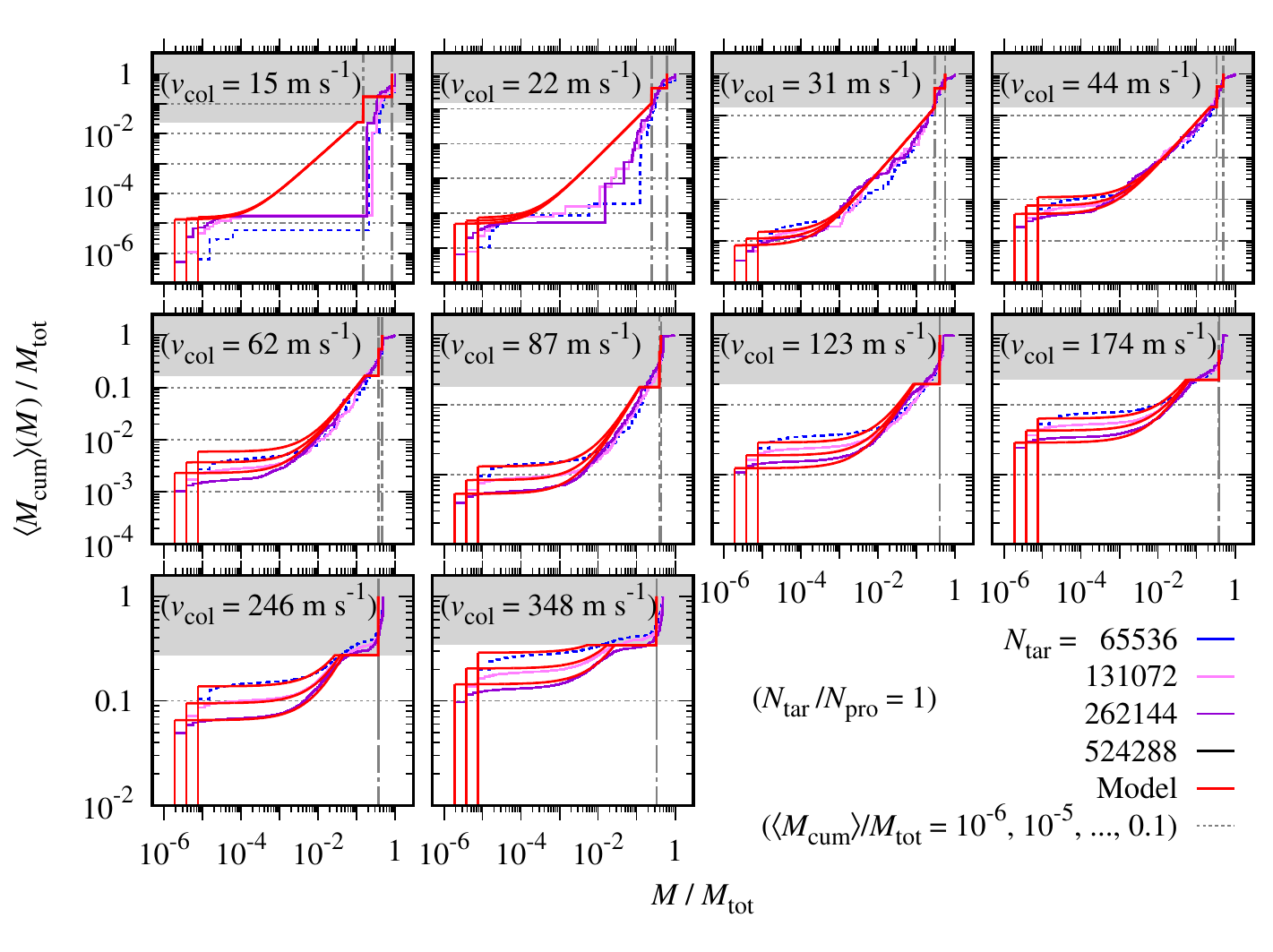}
  \caption{Same as Figure \ref{fig:NoNt_Ncum_Ntar_vcol_Ntop8}, but fitting relations derived from Equation (\ref{equ:fitting_NcumoNtot_model}) included (red solid lines). Vertical gray dashed-dotted and chain double-dashed lines denote $M = \langle M_{\rm lar} \rangle$ and $\langle M_{\rm 2nd} \rangle$, respectively. Gray shaded regions represent $\langle M_{\rm cum} \rangle (M) \ge \langle M_{\rm eje} \rangle$.}
  \label{fig:appendix_fitting_NoNt_Ncum_Ntar_vcol_Ntop1}
\end{figure}

\begin{figure}
  \plotone{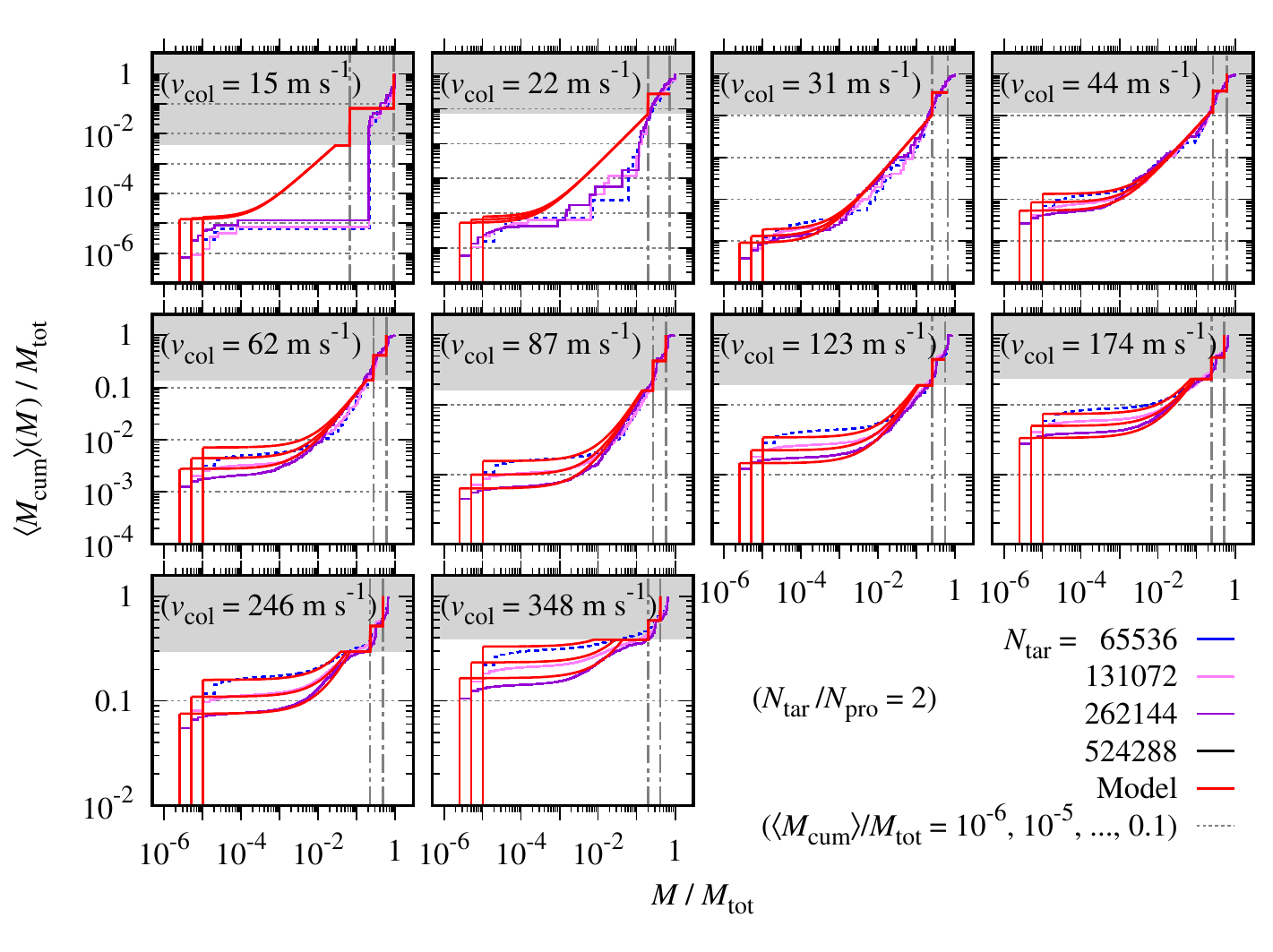}
  \caption{Same as Figure \ref{fig:appendix_fitting_NoNt_Ncum_Ntar_vcol_Ntop1}, but for $N_{\rm tar} / N_{\rm pro} = 2$.}
  \label{fig:appendix_fitting_NoNt_Ncum_Ntar_vcol_Ntop2}
\end{figure}

\begin{figure}
  \plotone{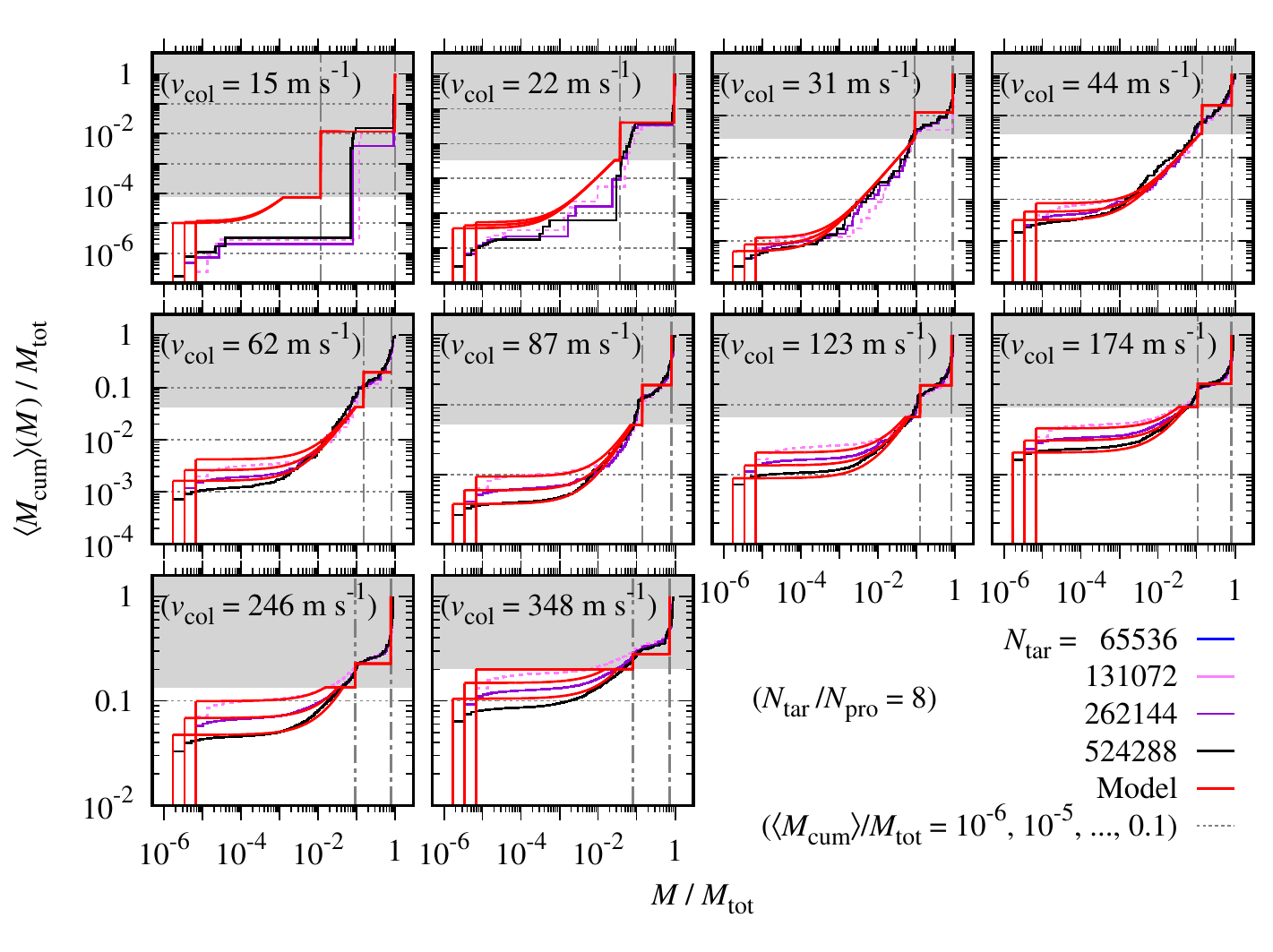}
  \caption{Same as Figure \ref{fig:appendix_fitting_NoNt_Ncum_Ntar_vcol_Ntop1}, but for $N_{\rm tar} / N_{\rm pro} = 8$.}
  \label{fig:appendix_fitting_NoNt_Ncum_Ntar_vcol_Ntop8}
\end{figure}

\begin{figure}
  \plotone{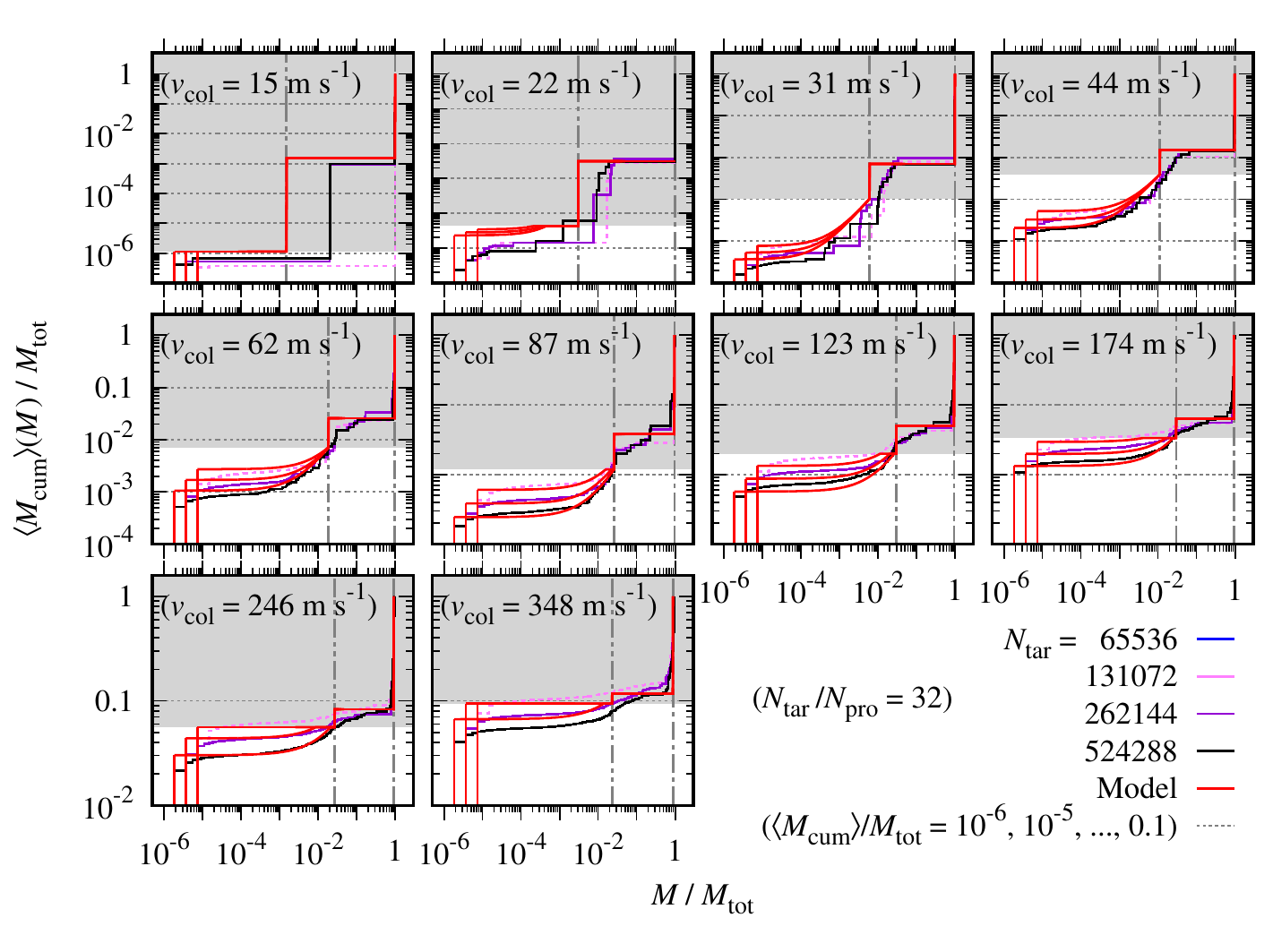}
  \caption{Same as Figure \ref{fig:appendix_fitting_NoNt_Ncum_Ntar_vcol_Ntop1}, but for $N_{\rm tar} / N_{\rm pro} = 32$.}
  \label{fig:appendix_fitting_NoNt_Ncum_Ntar_vcol_Ntop32}
\end{figure}

Figures \ref{fig:appendix_fitting_NoNt_Ncum_Ntar_vcol_Ntop1} to \ref{fig:appendix_fitting_NoNt_Ncum_Ntar_vcol_Ntop32} show the largest remnant mass, the second remnant mass, and the normalized cumulative mass distributions of the fragments, obtained by our simulations (lines except for red ones) and reproduced by our fragment model (red lines).
For high-velocity collisions of dust aggregates with the small total monomer number, all fragments except for the two large remnants are monomer fragments owing to the small total monomer number.

\bibliography{sample63}{}
\bibliographystyle{aasjournal}



\listofchanges

\end{document}